\documentclass{aa}  
\usepackage{graphicx}
\usepackage{txfonts}
\newcommand{\eps}{\varepsilon}
\newcommand{\epseri}{$\varepsilon$\,Eri }
\newcommand{\aumic}{AU\,Mic }
\title{Radial velocities with Machine Learning}
\author{M.~Perger}
\date{\today}
\begin{document}
\title{A Machine Learning approach for \\ correcting radial velocities using physical observables}
\author{
M.~Perger\inst{1,2}
G.~Anglada-Escudé\inst{1,2}
\and D.~Baroch\inst{1,2} 
\and M.~Lafarga\inst{3}
\and I.~Ribas\inst{1,2} 
\and J.\,C.~Morales\inst{1,2} 
\and E.~Herrero\inst{1,2} 
\and P.\,J.~Amado\inst{4}
\and J.\,R.\,Barnes\inst{5} 
\and J.\,A.~Caballero\inst{6}
\and S.\,V.~Jeffers\inst{7}
\and A.~Quirrenbach\inst{8}
\and A.~Reiners\inst{9} 
}
\institute{
\inst{1}Institut de Ciències de l’Espai (ICE, CSIC), Campus UAB, Carrer de Can Magrans s/n, 08193 Bellaterra, Spain\\    
\email{perger@ice.cat}\\
\inst{2}Institut d’Estudis Espacials de Catalunya (IEEC), 08034 Barcelona, Spain\\
\inst{3}Department of Physics, University of Warwick, Gibbet Hill Road, Coventry CV4 7AL, United Kingdom\\
\inst{4}Instituto de Astrof\'{\i}sica de Andaluc\'{\i}a (IAA-CSIC), Glorieta de la Astronom\'{\i}a s/n, 18008 Granada, Spain\\
\inst{5}School of Physical Sciences, The Open University, Walton Hall, Milton Keynes, MK7 6AA, UK\\
\inst{6}Centro de Astrobiolog\'{\i}a (CAB), CSIC-INTA, ESAC, Camino bajo del castillo s/n, 28692 Villanueva de la Ca\~nada, Madrid, Spain\\
\inst{7}Max Planck Institut f\"ur Sonnensystemforschung (MPS), Justus-von-Liebig-Weg 3, 37077 G\"ottingen, Germany\\
\inst{8}Landessternwarte, Zentrum f\"ur Astronomie der Universit\"at Heidelberg, K\"onigstuhl 12, 69117 Heidelberg, Germany\\
\inst{9}Institut f\"ur Astrophysik, Georg-August-Universit\"at G\"ottingen, Friedrich-Hund-Platz 1, 37077 G\"ottingen, Germany
}
\date{Received 29.09.2022 / Accepted 01.03.2023}
\abstract
% context heading (optional)
{Precision radial velocity (RV) measurements continue to be a key tool to detect and characterise extrasolar planets. While instrumental precision keeps improving, stellar activity remains a barrier to obtain reliable measurements below 1-2\,m\,s$^{-1}$ accuracy.}
{Using simulations and real data, we investigate the capabilities of a Deep Neural Network approach to produce activity free Doppler measurements of stars.}
{As case studies we use observations of two known stars ($\epsilon$\,Eridani and AU\,Microscopii, both with clear signals of activity induced Doppler variability. Synthetic observations using the {\tt starsim} code are generated for the observables (inputs) and the resulting Doppler signal (labels), and then they are used to train a Deep Neural Network algorithm to predict Doppler corrections. We identify a relatively simple architecture consisting of convolutional layers followed by fully connected layers that is adequate to the task. The indices investigated are mean line-profile parameters (width, bisector, contrast) and multi-band photometry.}
{We demonstrate that the RV-independent approach can drastically reduce spurious Doppler variability from known physical effects such as spots, rotation and convective blueshift. We identify the combinations of activity indices with most predictive power. When applied to real observations, we observe a good match of the correction with the observed variability, but we also find that the noise reduction is not as good as in the simulations, probably due to the lack of detail in the simulated physics.}
{We demonstrate that a model-driven machine learning approach is sufficient to clean Doppler signals from activity induced variability for well known physical effects. There are dozens of known activity related observables whose inversion power remains unexplored indicating that the use of additional indicators, more complete models, and more observations with optimised sampling strategies can lead to significant improvements in our detrending capabilities on new and existing data sets.}
\maketitle
%________________________________________________________________
%
\section{Introduction}

The majority of exoplanet discoveries in the last two decades have used two methods. One is the transit method monitoring tens of thousands of stars simultaneously, and most effectively done from space with instruments such as Kepler \citep{2010Sci...327..977B} and TESS \citep[Transiting Exoplanet Survey Satellite;][]{2015JATIS...1a4003R}. The other one is the Doppler technique, which is conducted using high-resolution spectrographs such as HIRES \citep[HIgh Resolution Échelle Spectrograph;][]{1994SPIE.2198..362V}, HARPS \citep[High Accuracy Radial velocity Planet Searcher;][]{2003Msngr.114...20M}, HARPS-N \citep{2012SPIE.8446E..1VC}, CARMENES \citep[Calar Alto high-Resolution search for M dwarfs with Exoearths with Near-infrared and optical Échelle Spectrograph; ][]{2020SPIE11447E..3CQ}, or ESPRESSO \citep[Échelle SPectrograph for Rocky Exoplanets and Stable Spectroscopic Observations;][]{2021A&A...645A..96P}. Whereas the field of exoplanet science is now progressing from the mere discovery of planetary companions to the analysis of atmospheres of already known, transiting planets, those surveys are still collecting new data providing complementary insight into planet formation and evolution. Significant challenges remain in the detection of small terrestrial planets around Sun-like stars mainly due to spurious effects caused by stellar variability.

There are large datasets, both collected from photometric and spectroscopic observations, which may have the precision to detect objects of high interest, but remain unexploited. In particular, radial velocities (RVs) are extracted from the high-resolution spectra by measuring the wavelength shift of the different lines present in the spectra compared to a laboratory reference standard or to a template created from the stellar spectra themselves. Shifts can be caused by motion of the stellar center of mass caused by a planet, or they can be induced by the stellar activity. There are various effects influencing those measurements. \citet{2012Natur.491..207D} give a comprehensive overview on the topic. The effects introduced by stellar surface phenomena, such as dark spots or bright faculae, induce apparent RV wobbles with typical time scales of rotation periods of stars, active region life-times, and stellar activity cycles, making it extremely difficult to disentangle such effects from true Doppler signals from planets, which tend to have comparable orbital periods. Most effective \textit{inversion} strategies involve the joint fit of the so-called activity indices with the RVs using a handful of shared parameters in Gaussian processes as discussed in \citet{2015MNRAS.452.2745S}, which still has the drawback of reducing the information content of the time series, and potentially filtering out potential signals of interest.

Currently RV precisions of 1\, m\,s$^{-1}$ are regularly achieved. But this precision rarely translates into accurate measurements of true stellar motion due to the effects of stellar activity. This spurious variability is estimated to be about 2 to 4\,m\,s$^{-1}$ \citep{2017A&A...598A..26P} for quiet old stars, to hundreds of m\,s$^{-1}$ for young and active stars. To account for those effects, many empirical techniques have been proposed and applied with limited success. This includes a simple fit of one or two sinusoids accounting for the stellar activity \citep[][often called pre-whitening]{2015MNRAS.452.2745S}; in-depth analysis of additional activity indices, photometry and their correlations with observed RV signals; and the application of more sophisticated noise models in Bayesian frameworks, such as Gaussian processes \citep{2019A&A...624A.123P, 2021A&A...645A..58P}. In recent times, as for many fields in astrophysics and science in general, machine learning techniques are being considered to solve similar problems. \citet{2018AJ....155...94S} and \citet{2019AJ....157..169D} developed and applied a method for classifying potential planet signals using machine learning algorithms in the Kepler/K2 data. \citet{2020arXiv201100003D}, on the other hand, use the cross-correlation functions (CCFs) created by the binary-mask technique used to extract the RVs from HARPS-N spectra of the Sun \citep{2019MNRAS.487.1082C} and from SOAP 2.0 \citep{2014ApJ...796..132D} synthetic data, and then train a neural network (NN) algorithm to distinguish Doppler shifts due to companions from activity induced shifts based on the information in the cross-correlation profile.

In this article, we investigate whether a machine learning framework trained using synthetic (but realistic and complex) observables is able to produce usable RV corrections without using the information of the Doppler shifts themselves. We present the set up and assumptions of our study in Sect.\,\ref{sec:datasetup}, including the different activity indices studied, two stars used as study-cases, the models and software used to generate synthetic time-series ({\tt starsim}), and the methodologies and machine learning framework that we have developed to address the problem. The results and discussions are given in Sect.\,\ref{sec:results}, and we provide conclusions to our study in Sect.\,\ref{sec:conclusions}.

\section{Set up} \label{sec:datasetup}

\subsection{Tracers for stellar activity} \label{subsec:activityindex}

In the Doppler technique, the stellar spectrum is measured with high spectral resolution ($\mathcal{R} > 50\,000$). The Doppler shift of the spectrum is then obtained by cross-correlating the observed spectrum with a mask \citep[cross-correlation technique as in][]{2012SPIE.8446E..1VC, 2020A&A...636A..36L} or a template built from the same observations \citep{2012ApJS..200...15A, 2018A&A...609A..12Z}. The high resolution of the spectra also allows for simultaneous measurement of spectral characteristics other than the RV. In addition to this, contemporaneous photometric observations can be used to infer the coverage of spots on the star, putting constraints on the expected RV variability. When possible, this photometric monitoring is done in various filter bands (or colours) in order to determine not only the spot coverage but also properties of the spots such as their effective temperatures \citep{2020A&A...641A..82R}.

The process of measuring precision Doppler shifts with the cross-correlation method consists of consolidating the information of hundreds (thousands) of spectral lines in a cross-correlation profile obtained by performing the correlation of the observed spectra with a predefined weighted mask (list of wavelengths where spectral lines should appear at zero Doppler shift). The resulting CCF profile is equivalent to the \textit{mean} shape of the spectral lines for each observation. The computation of this CCF profile is typically produced by instrument pipelines such as the Data Reduction Systems (DRS) of the instruments HARPS, HARPS-N, and ESPRESSO, and the {\tt raccoon} code \citep[Radial velocities and Activity indicators from Cross-COrrelatiON with masks; ][]{2020A&A...636A..36L} developed in the context of the CARMENES project. In this study we use {\tt raccoon} to obtain Doppler shifts and precision measurements of the shape parameters of the CCF. 

The RV is measured by fitting a Gaussian function and measuring its mean. Other useful measurements can be obtained from the fitted Gaussian profile, such as the full-width-at-half maximum (or FWHM, in\,m\,s$^{-1}$) and the relative depth -or contrast- relative to the continuum (CON, in \%). As an indicator of possible changes in the symmetry of the CCF profile, the Bisector Inverse Slope (BIS, in m\,s$^{-1}$) is usually measured. As abundantly discussed in the literature, all these \textit{indices} (FWHM, CON, and BIS, among others) are known to be affected by the presence of spots and other physical effects that are also responsible for the observed spurious Doppler shifts. Therefore the information needed to produce a \textit{clean} RV measurement could be, in principle, available in the same observation through the indices. This production of clean RVs from other spectroscopic indices (or \emph{inversion} hereafter) has been the topic of many research efforts during the last decade achieving only partial success. 

\subsection{Observational data} \label{subsec:examplestars}

In order to base our experiments with synthetic data on realistic observations, we use measurements of two well studied and representative stars: the very nearby K2-dwarf $\varepsilon$\,Eridani (Gliese\,144; \epseri hereafter), and the very young M1-dwarf AU\,Microscopii (Gliese\,803; \aumic hereafter). They are both quite active and show strong stellar contributions to the RVs as well as to all CCF indices. In Table\,\ref{tab:teststars}, we provide the parameters of these test stars as found in the literature. Both datasets were chosen for having high signal to noise spectra (so observations would not be limited by photon noise), and their observing cadence was quite regular over a relatively short time interval (2--3 months).

\begin{table*}
\caption{Literature and {\tt starsim} parameters for $\varepsilon$\,Eridani, and AU\,Microscopii, and their planetary companions.} \label{tab:teststars}
\scriptsize
\centering                         
\begin{tabular}{l l | c c c | c c c}     
\hline\hline                 
\noalign{\smallskip}
parameter & unit & \multicolumn{3}{c|}{$\varepsilon$\,Eridani} & \multicolumn{3}{c}{AU\,Microscopii} \\
 & & literature value  & Ref. & {\tt starsim} & literature value  & Ref. & {\tt starsim} \\ 
\hline
\noalign{\smallskip}
Right ascension RA & h:m:s & 03:32:55.84$\pm$0.12 & (1) & ... & 20:45:09.532$\pm$0.017 & (1) & ... \\
Declination Dec     & º:':'' & $-$09:27:29.739$\pm$0.093 & (1) & ... & $-$31:20:27.238$\pm$0.011 & (1) & ... \\
age & Myr & 400 - 800 & (2) & ... & 22$\pm$3 & (4) & ... \\
distance & pc & 3.216$\pm$0.002 & (1) & ... & 39.714$\pm$0.002 & (1) & ... \\
$\mu_{\rm \alpha}$ & mas\,a$^{-1}$ & $-$974.76$\pm$0.16 & (1) & ... & 281.319$\pm$0.022 & (1) & ... \\
$\mu_{\rm \delta}$ & mas\,a$^{-1}$ & 20.88$\pm$0.12 & (1) & ... & $-$360.148$\pm$0.019 & (1) & ... \\
$v \sin{\rm i}$ & km\,s$^{-1}$ & 2.4$\pm$0.5 & (3) & ... & 9.3$\pm$1.2 & (5) & ... \\
magnitude $G$ & mag & 3.4658$\pm$0.0031 & (1) & ... & 7.8434$\pm$0.0029 & (1) & ... \\ \hline
\noalign{\smallskip}
spectral type SpT & ... & K2.0V  & (6) & HARPS K5 mask & M1.0V  & (17) & CARMENES M1.5 mask\\
stellar mass $M_*$v& M$_{\odot}$ & 0.82$\pm$0.05 & (7)  & 0.8 - 0.9 & 0.50$\pm$0.03 & (4) & 0.45 - 0.55   \\
 &             & 0.847$\pm$0.042 & (8) &              &               &   &       \\   
% &             & 0.871$\pm$0.003 & (9) &              &               &   &       \\    
stellar radius $R_*$ & R$_{\odot}$ & 0.74$\pm$0.01 & (7)  & 0.65 - 0.75 & 0.75$\pm$0.03 & (4) & 0.7 - 0.8   \\
 &             & 0.702$\pm$0.035 & (8) &              &               &   &       \\
% &             & 0.911$\pm$0.007 & (9) &              &               &   &       \\          
effective temperature $T_{\rm eff}$ & K & 5076$\pm$30   & (10) & 4900 - 5300 & 3700$\pm$100   & (4) & 3550 - 3850  \\
surface gravity $\log{\rm g}$ & dex  & 4.30$\pm$0.08 & (11) & 4.5 (fixed) & 4.39$\pm$0.03 & (1) & 4.5 (fixed)  \\
metallicity $[Fe/H]$ & ...          & $-$0.13$\pm$0.04 & (12) & 0.0 (fixed) & $-$0.12$\pm$0.11 & (18) & 0.0 (fixed) \\ 
inclination $i$ & $^{\circ}$        & 60  & (13) & 30 - 90 & 90  & (19) & 60 - 90  \\ 
rotation period $P_{\rm rot}$ & day & 11.2 & (3) & 11 - 12.5 & 4.863$\pm$0.010 & (20) & 4.8 - 4.9 \\ 
convective blueshift $CB^{*}$ & $CB_{\odot}$ & $\sim$0.3  & (15,16) & 0.0 - 0.5  & $\sim$0.0  & (15,16) & 0.0 - 0.5  \\
differential rotation $d\Omega^{**}$ & $d\Omega_{\odot}$        & 1.3  & (3) & 0.1 - 4 & 3.0  & (21) & 0.5 - 5 \\ 
spot temperature difference $\Delta T$ & K & 1080$\pm$670 & (14) & 250 - 1000 & 660$\pm$570 & (14) & 250 - 1000 \\
\hline \noalign{\smallskip}
minimum mass $M \sin{\rm i}$ & M$_{\oplus}$ & 210 & (22) & ... & 20; 11  & (23) & ... \\
orbital period & day & 2671 & (22) & ... & 8.46; 18.86 & (23) & ... \\
RV semi-amplitude & m\,s$^{-1}$ & $\sim$11 &  ... & ... & $\sim$6; $\sim$8    & ... & ... \\
\noalign{\smallskip} \hline               
\end{tabular}
\tablefoot{
\tablefoottext{1}{\citet{2020yCat.1350....0G}}, 
\tablefoottext{2}{\citet{2015A&A...574A.120J}}, 
\tablefoottext{3}{\citet{2007AN....328.1037F}}, 
\tablefoottext{4}{\citet{2020Natur.582..497P}}, 
\tablefoottext{5}{\citet{2006A&A...460..695T}}, 
\tablefoottext{6}{\citet{1989ApJS...71..245K}}, 
\tablefoottext{7}{\citet{2012ApJ...744..138B}},
\tablefoottext{8}{\citet{2019A&A...623A..72K}},
\tablefoottext{9}{\citet{2021A&A...646A..77G}},
\tablefoottext{10}{\citet{2015A&A...582A..49H}},
\tablefoottext{11}{\citet{2010MNRAS.403.1368G}}, 
\tablefoottext{12}{\citet{2004A&A...415.1153S}}, 
\tablefoottext{13}{\citet{2002ApJ...578L.149Q}}
\tablefoottext{14}{following \citet{2021ApJ...907...89H}},
\tablefoottext{15}{\citet{2021A&A...654A.168L}}, 
\tablefoottext{16}{\citet{2016csss.confE.104M}}, 
\tablefoottext{17}{\citet{1989ApJS...71..245K}},   
\tablefoottext{18}{\citet{2014MNRAS.443.2561G}}, 
\tablefoottext{19}{\citet{2005Natur.435.1067K}}
\tablefoottext{20}{\citet{2005AJ....129.1008K}},  
\tablefoottext{21}{\citet{2021MNRAS.502..188K}}, 
\tablefoottext{22}{\citet{2021AJ....162..181L}},
\tablefoottext{23}{\citet{2022MNRAS.512.3060Z}},\\
\tablefoottext{*}{in units of the solar convective blueshift CB$_{\odot} \sim -325$\,m\,s$^{-1}$ \citep[e.g.,][]{2018A&A...611A...4L, 2019A&A...622A..34S}},\\
\tablefoottext{**}{in units of the solar differential rotation $d\Omega_{\odot} = 3.15\,^{\circ}\,{\rm day}^{-1}$, assuming $P_{\odot,\rm min} \sim  24.47\,{\rm d}$, $P_{\odot,\rm max} \sim 31.14$\,d, and $d\Omega_{\odot} = 360\,^{\circ} (P_{\odot,\rm min}^{-1} - P_{\odot,\rm max}^{-1}$) \citep{2000SoPh..191...47B}}.
}
\end{table*}

\epseri is a young, nearby star with a rotation period of about 11.2\,d \citep{2007AN....328.1037F}. It has a debris disk inclined by 30\,$^{\circ}$ \citep{2002ApJ...578L.149Q} with respect to the sky plane, and at least one planetary companion of at least 0.7\,$M_{\rm Jup}$ in a 7-year period orbit \citep{1988ApJ...331..902C} causing a Doppler signal with a RV semi-amplitude of about 11\,m\,s$^{-1}$. For this study, we use 204 publicly available spectra obtained with HARPS ($3\,750\,\AA{} < \lambda <6\,900\,\AA{}$, $\mathcal{R}=115\,000$) obtained on 66 quasi-consecutive nights between 5 Oct 2019 (${\rm BJD} =2458762$\,d) and 1 Jan 2020 (${\rm BJD} = 2458850$\,d) with a total baseline of 88\,day\footnote{Programme IDs: 0104.C-0863, 072.C-0488, 072.C-0513, 073.C-0784, 074.C-0012, 076.C-0878, 077.C-0530, 078.C-0833, 079.C-0681, 192.C-0852, 60.A-9036 secured as part of the Red Dots ESO programme.}. All observations within a night were consolidated in night average values to produce the 66 epochs that we will use in all that follows.

AU\,Mic, on the other hand, is a young star located at a distance of 9.71\,pc \citep{2020Natur.582..497P}. It rotates with a period of 4.9\,d \citep{2005AJ....129.1008K} and shows strong evidence of differential rotation \citep[$d \Omega = 9.57 \pm 0.52\,^{\circ}\,{\rm day}^{-1}$,][]{2021MNRAS.502..188K}. Its debris disk is seen edge-on \citep{2005Natur.435.1067K}. Since planets are supposed to form in such disks, it could host transiting planets as well. Indeed, \citet{2020Natur.582..497P} and follow-up publications reported the detection of (at least) two transiting planetary companions with orbital periods of 8.5\,d, and 18.9\,d, and minimum masses of about 20 and 11\,M$_{\oplus}$, respectively, inducing possible RV semi-amplitudes of 6, and 8\,m\,s$^{-1}$. In this study we use 75 CARMENES VIS spectra ($5\,200\,\AA{} < \lambda <9\,600\,\AA{}$, $\mathcal{R}=94\,600$) obtained for this star \citep[available from CARMENES DR1,][]{Ribas2023}, which were observed quite regularly on 37 nights from 14 Jul 2019 (${\rm BJD} =2458679$\,d) to 9 Oct 2019 (${\rm BJD} =2458766$\,d) with a total baseline of 87\,days.After consolidating the data in night averages, we reach an average time spacing of 2.5\,days between consecutive observations. 

Our early experiments were run using DRS instrument pipeline data-products for the real data, and {\tt raccoon} generated measurements on synthetic data. The mismatches were very obvious in all cases and no correction close to what was observed could be obtained from them. This indicated that a high level of consistency was needed in all the algorithms producing observables in the simulations and the real data-products. In order to be fully consistent, we use the {\tt raccoon} code to retrieve RV, FWHM, BIS, and CON from the CCF using the binary masks as indicated in Table\,\ref{tab:teststars}, and then adapted the codes in {\tt starsim} to be fully consistent with them. 

\begin{figure*}
\centering
\includegraphics[width=\textwidth]{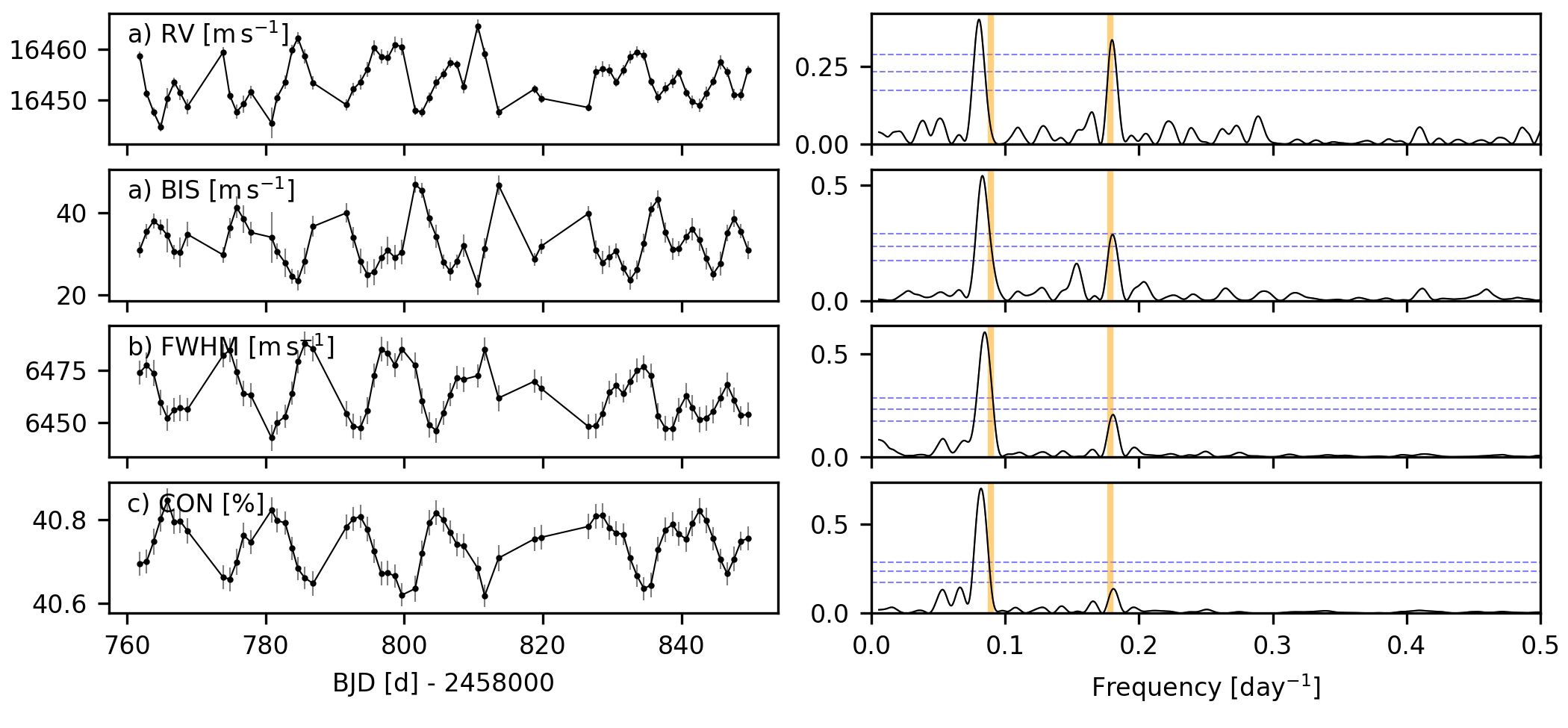}
\caption{Nightly binned time series (left panels) and their periodograms (right panels) of \epseri extracted with {\tt raccoon}. From top to bottom those are RV, BIS, FWHM, and CON. Dashed blue horizontal lines indicate the FAP levels of 0.1, 1, and 10\,\%, and the vertical orange lines indicate the frequency of the rotation period of 11.2~d and its second harmonic.} \label{fig:timeseries1}
\end{figure*}

\begin{figure*}
\centering
\includegraphics[width=\textwidth]{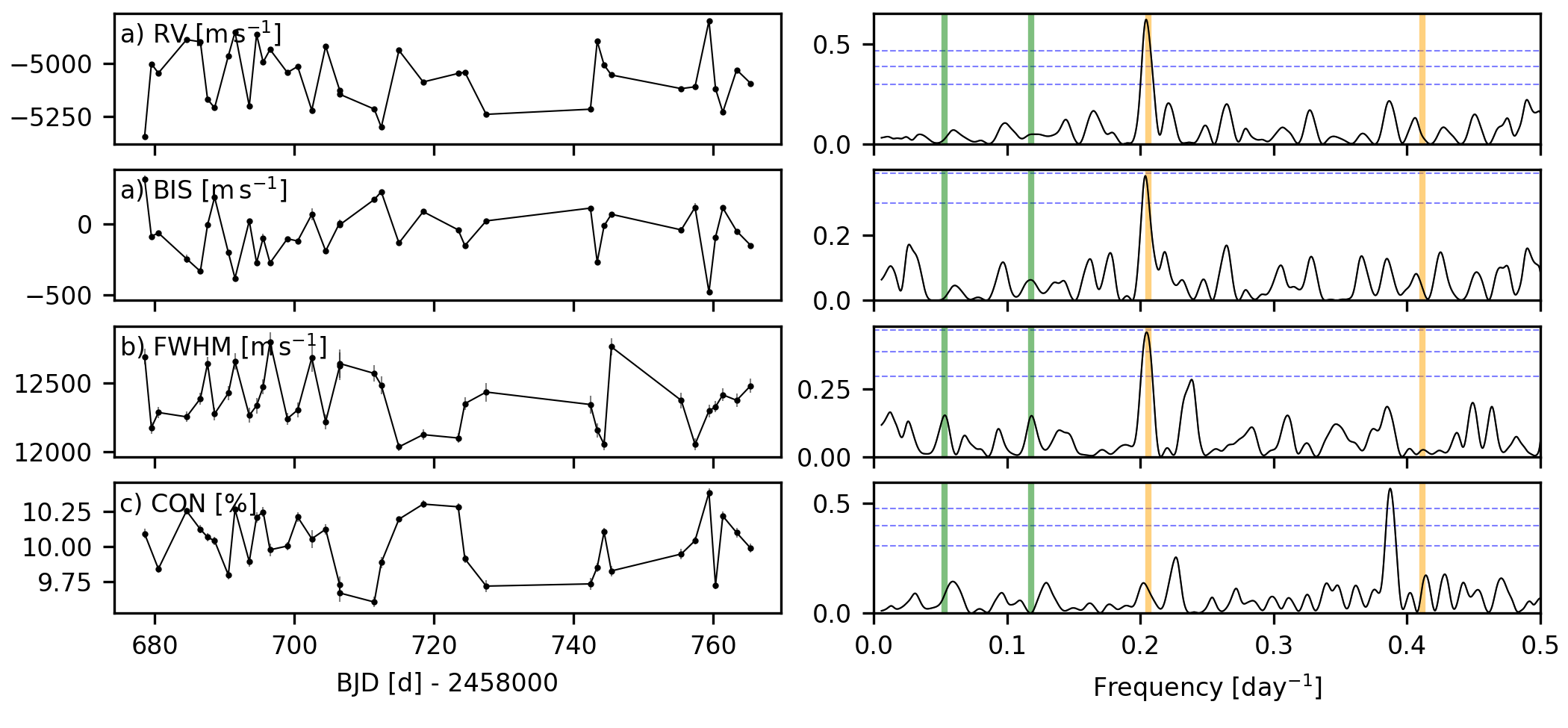}
\caption{Nightly binned time series (left panels) and their periodograms (right panels) of \aumic extracted with {\tt raccoon}. From top to bottom those are RV, BIS, FWHM, and CON. For the latter, one obvious outlier is not in this diagram. Dashed blue horizontal lines indicate the FAP levels of 0.1, 1, and 10\,\%, and the vertical orange lines indicate the frequency of the rotation period of 4.9~d and its second harmonic. The vertical green lines mark the periods of the two planetary companions.} \label{fig:timeseries2}
\end{figure*}

For the analyses that follow, it is important to separate what we consider random errors (photon noise, instrumental and other effects generating uncorrelated errors) from activity induced variability (which has time-scales of days, due to the rotation period and its harmonics), especially for the real datasets. To obtain an estimate of the random errors for all the real time-series, (RV, but also the activity indices), we perform a pre-whitening procedure using sequential fitting of sinusoids using Generalised Lomb-Scargle periodograms \citep[GLS,][]{2009A&A...496..577Z}. When there is no signal left below a 1\,\% false alarm probability (FAP), we assume that the remaining variability are these random errors, although they might contain some correlated stellar jitter below the noise level. The standard deviation of these random errors in all the series will later be used to simulate synthetic datasets with \emph{realistic} observational noise. Table\,\ref{table:timeseries} shows the residual RMS from this procedure, and the ratio between the original RMS and the residual RMS after the pre-whitening process, expressed as a percentage. This ratio between the residual RMS compared to the original RMS expressed as a percentile is what we call \emph{Relative Residual Error} (or RRE), and it will be used in many instances throughout the paper for different time-series. We rather use this value because we compare our results to the relative RV change and not to its absolute true value (which would be the accuracy). A depiction of the time-series (RV and spectroscopic indices) as obtained from the observations is shown in Figs.\,\ref{fig:timeseries1} ($\varepsilon$\,Eri) and \ref{fig:timeseries2} (AU\,Mic). The data in the figures can be found in Tables\,\ref{table:HNdata} and \ref{table:CARMdata}.

\begin{table*}
\small
\caption{Time-series data characteristics of 66 nightly binned data points of \epseri using HARPS (top), and of 37 nightly binned data points from \aumic using CARMENES VIS (bottom), both extracted by {\tt raccoon}
.} \label{table:timeseries} 
\centering                         
\begin{tabular}{llccccl}     
\hline\hline \noalign{\smallskip} 
data & unit &  mean & RMS  & uncertainty & RRE & significant periods \\
&  &  [unit] & [unit] & [unit] & [\%] & [d] \\
\hline \noalign{\smallskip}
RV  & m\,s$^{-1}$ & 16453.68 & 4.36 & 1.19 & 39 & 12.44; 5.56 \\                  
BIS & m\,s$^{-1}$ & 32.48 & 5.56 & 2.44 & 17 & 12.06; 5.56; 10.60  \\
FWHM & m\,s$^{-1}$ & 6463.83 & 12.01 & 5.72 & 23 & 11.81; 5.54; 175.43; 12.81; 10.32 \\
CON & \% & 40.739 & 0.058 & 0.029 & 25 & 12.18; 5.54; 175.43; 10.63; 12.62  \\     
\noalign{\smallskip} \hline \noalign{\smallskip}
RV  & m\,s$^{-1}$ & $-$5064.84 & 132.26 & 6.76 & 25 & 4.90; 2.59 \\                  
BIS & m\,s$^{-1}$ & $-$61.75 & 172.04 & 17.11 & 28 & 4.90; 37.72; 2.59; 5.74; 2.45 \\
FWHM & m\,s$^{-1}$ & 12380.18 & 202.05 & 54.23 & 58 & 4.89; 4.19  \\          
CON$^*$ & \% & 10.013 & 0.204 & 0.035 & 71 & 2.58  \\     
\noalign{\smallskip} \hline          
\end{tabular}
\tablefoot{Relative residual error (RRE) is the ratio between the original RMS versus the resulting one obtained after applying pre-whitening (aka. until no signal with FAP$<1$\% is present in the data)\\
*: statistics without outlier}
\end{table*}

The signal of the stellar rotation period is present in all time series of \epseri. Probably due to the differential rotation of the star, the rotation period as given in the literature manifests itself more strongly in its second harmonic at 5.6\,days. By using the pre-whitening procedure described earlier, we achieve RREs between 17 to 25\% in all indices. 

For \aumic we find the rotation period in all the data sets except the CON series. As also discussed later, the measurement of the CON series might not be useful for this kind of target and study. For the other indices the achieved RRE is about 30\% to 60\%. 

\begin{figure}
\centering
\includegraphics[width=9cm]{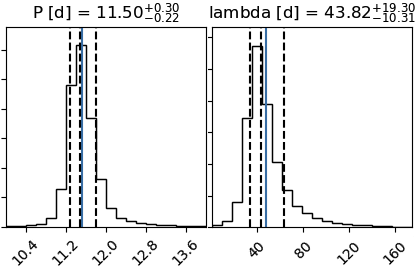}
\includegraphics[width=9cm]{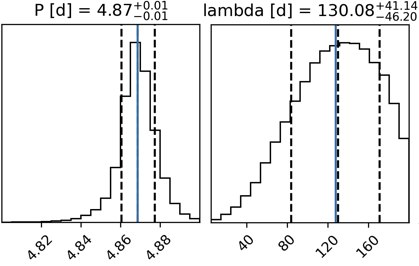}
\caption{Histogram of the MCMC posterior distributions for hyperparameters $P$ (left panels) and $\lambda$ (right panels) of Gaussian processes applied to \epseri (top panels) and \aumic (bottom panels) RV data using the QPC kernel.} \label{fig:QPC}
\end{figure}

Next we perform a quick analysis applying a Gaussian Process (GP) using the {\tt george} package \citep{2015ITPAM..38..252A} and the quasi-periodic and cosine (QPC) kernel \citep{2021A&A...645A..58P}, and exploring the parameter space with a Monte-Carlo Markow Chain (MCMC) using {\tt emcee} \citep{2013PASP..125..306F}. We show in Fig.\,\ref{fig:QPC} the posterior distributions for the period hyperparameters $P_{\rm \eps Eri}=11.50^{+0.30}_{-0.22}$\,d and $P_{\rm \aumic}=4.87 \pm 0.01$\,d, and $\lambda=44^{+19}_{-10}$\,d and $130^{+41}_{-46}$\,d , where $\lambda$ is the characteristic decay time of the correlations (related to the average time it takes for a spot to change substantially in size or location, or even disappear) for the RVs of \epseri and AU\,Mic, respectively.

\subsection{Synthetic data} \label{subsec:starsim}

We use the {\tt starsim} code \citep{2016A&A...586A.131H} to model time-series data for a spotted rotating star with the characteristics of our two tests stars \epseri and AU\,Mic. The current version of {\tt starsim} \footnote{\url{https://github.com/dbarochlopez/starsim}} (Baroch et al. in prep.) is able to produce synthetic photometry for a predefined filter from low-resolution model spectra for our study, we produce light curves in Johnsons $V$, $R$, and $I$ filters. CCFs at each observation epoch are also generated from high-resolution Phoenix spectra \citep{2013A&A...553A...6H} of a predefined wavelength range, from which RVs, and the CCF activity indices are then measured. The calculations make use of important stellar parameters, geometric effects, and of the distribution of \emph{magnetically active regions} (spots hereafter) on the surface of the star. One simulation consists of a list of epochs and their corresponding RV, FWHM, BIS, CON, and $V$, $B$ and $I$ photometric bands, all assumed to be simultaneous and without errors. 

Preliminary experiments (not described here for brevity) showed that constraining the simulations to the known physical parameters of the star is rather important to improve the inversion capabilities of the algorithms used. This makes sense as NNs are known to be good tools for interpolation as long as the training set densely populates the region of the parameter space of interest. The priors of the most important stellar variables as adopted for the creation of the synthetic data sets of the two test stars are shown in Table\,\ref{tab:teststars}. For mass, radius, effective temperature, inclination of the rotation axes, and the rotation period we use the published values in their error limits. In the case of $\varepsilon$\,Eri, we adapt the range for the latter parameter in order to include the larger periodicities found by our pre-whitening as shown in Table\,\ref{table:timeseries}. We choose further fixed values for metallicity and surface gravity. In the algorithm, the convective blueshift is given in units of the solar shift of $-325$\,m\,s$^{-1}$ \citep[e.g.,][]{2018A&A...611A...4L, 2019A&A...622A..34S}, and the differential rotation in units of the solar value $d\Omega_{\odot}=3.15\,^{\circ}\,{\rm day}^{-1}$ \citep[following][]{2000SoPh..191...47B}. Furthermore, the binary masks of the corresponding stellar type and wavelength range were used to calculate the CCFs. In each simulation, the physical properties of the star are randomly initialised within these priors, thus implicitly accounting for the uncertainty in the basic properties of the star at simulation level (training and test sets).

{\tt starsim} also requires to specify the distribution of the spots on the stellar surface (assumed circular) in what we call a \emph{spot map}. The contribution of a spot in a simulation is taken into account when producing a CCF integrated over the stellar disk (spots have a different CCF than the photosphere because of lower temperature and different spectral features). For the spots we also use Phoenix spectra of lower temperature as given by the $\Delta T$ parameter in Table\,\ref{tab:teststars}, and a modified bisector following what is observed on the Sun \citep{2016A&A...586A.131H}. The priors of these parameter are specified by following empirical studies on a variety of stellar types in \citet{2021ApJ...907...89H}. In the input \emph{spot map} each spot is described by five parameters: time of appearance, lifetime, longitude, co-latitude and radius. Taking into account that the contribution from stellar activity is clearly visible in all sets at all times (no flat regions in the observational data), and to match both absolute value and variation of each time series, we use randomly 25 to 40 dark spots to ensure that at least one to two spots are visible at any observing time. All spot map parameters are also generated randomly following a uniform distribution: radii (2.5 to 4\,$^{\circ}$), lifetime $\tau$ (5 to 100\,days), time of appearance ($t_0-\tau$ to $t_0 + 100$\,d), co-latitude (0 to 180\,$^{\circ}$), and longitude (0 to 360\,$^{\circ}$). The values adopted try to encompass the intervals of the most likely values from these parameters following the GP analysis from the previous section.

To investigate the importance of sampling in the ability to correct for stellar activity, we generate simulations following five different time samplings with {\tt starsim}
\begin{itemize}
    \item {\bf uniform} consisting of 100 data points uniformly distributed over 100\,days;
    \item {\bf random1} consisting of 100 data points randomly distributed over the same time baseline 
    \item {\bf gap1 } consisting of 60 randomly distributed data points from time 0\,d to 50\,d; and 40 data points from time 70\,d to 100\,d, thereby creating an artificial visibility gap as we often encounter it in actual astronomical campaigns
    \item {\bf random2} which uses 100 randomly chosen out of the 300 data points in \emph{uniform}, \emph{random1} and \emph{gap1}
    \item{\bf gap2} which has 40 data points in the first 30\,days, and 60 in the last 45\,days of the 300 data points, leaving open another gap.
\end{itemize}
\noindent where \emph{random2} and \emph{gap2} are generated from subsets of the previous samplings to save simulation time (most computationally intensive step as discussed later). The 100\,day time interval was chosen to cover the rotation period of both test stars a sufficient number of times. We then generate 600\,000 and 1\,000\,000 simulations for \epseri and AU\,Mic, respectively, for each of these time-samplings. 

We checked for possible linear relations between the synthetic RVs and the activity indices. Applying those relations, we reach RREs for $V$, $I$, BIS, FWHM, and CON indices of 99\%, 99\%, 79\%, 98\%, and 98\% for \epseri, and 99\%, 99\%, 73\%, 86\%, and 95\% for \aumic, respectively. If we apply a linear fit to all six datasets, we achieve RREs of 77\% for \epseri and 72\% for \aumic, respectively, showing the importance of a more sophisticated and RV-independent approach for our synthetic data.

\subsection{Neural Network Architectures} \label{subsec:neuralnetwork}

Finding the right format for the inputs, and defining an adequate NN architecture, are some of the major challenges when applying NN approaches to new types of data. Neural networks with several fully connected layers have a huge number of parameters, thus enabling a fit to almost anything. However, being so general, these are slow to train (and may not even converge at all) and often lead to over-fitting (i.e. the NN is able to reproduce the training set, but performs poorly on a test sample). For an NN to be able to make non-linear predictions, more than two layers are needed. Since the data we are analysing contain strong correlations between neighboring points, and these correlations encode the physics, we propose an NN composed of a small number of convolutional layers first, followed by three dense fully connected layers. This architecture is similar to those used in pattern recognition in images, which also looks for patterns in the spatial domain to identify features in an image. We use the open-source python package {\tt PyTorch}\footnote{{\tt PyTorch} is a trademark of Facebook, Inc.}. We also run experiments with Keras/Tensorflow\footnote{Maintained by Google}, which is the other most popular platform for NN development. Neither results nor performance depended much on the choice of the framework, so all results presented here were produced using {\tt PyTorch}. In all cases, we installed the GPU enabled versions of the libraries to take full advantage of our hardware capabilities. The codes were developed in python. Both the {\tt starsim} simulations and the training of the NN were executed on two identical dedicated machines (Intel Core i7-10700, which contains 8 CPU cores, and equipped with a GPU-NVIDIA GeForce RTX 2060 SUPER each).

An input sample consists of $N_{i}$ time-series of indices (or channels) corresponding to the activity indices used for each test. The target (also called label) of each input sample is the corresponding RV time-series of the corresponding simulation. For example, for a synthetic set with 100 epochs, a NN network trained to produce RVs from FWHM and BIS data, would have two channels (FWHM and BIS) of 100 datapoints each as input, and one output list with 100 RV datapoints. All time-series are mean subtracted and normalised by the full ensemble standard deviation for that index (or label). Thereby we conserve the information of the relative strengths of the variations of the indices. 

The input layer is then an array of $1 \times 100 \times N_{i}$ neurons. The first two layers are convolutional layers with small kernel sizes (typically 3 elements only), and a rather large number of channels (typically 32). This means that in the first convolutional layer, the information of each point and neighbors will be taken into account at the layer output. The second convolutional layer repeats the same operation, meaning that, effectively, the NN will take into account the value of each point and the $\pm$2 closest neighbors. Tests using kernels with larger sizes or the inclusion of more convolutional layers did not improve the performance of the NN significantly. The outputs of all neurons of the second convolutional layer are then arranged in a single list (so-called flatten operation), and this is then fed to three fully connected layers with 300 neurons each (also called dense layers). The output layer is a fully connected layer with 100 neurons. Except for the output layer (which has a linear function as its response), the response functions used in all other layers were the so-called ReLU (Rectified Linear Unit; $f(x)= {\rm max}(0,x))$ plus a constant offset (also called bias). We tested alternative functions (such as sigmoids), finding similar performances but at the cost of more training iterations and longer training time per iteration. As a cost function, we use the standard deviation of the difference between the simulated RVs (label) and the RV predictions by the NN (output). Training a NN consists of finding a combination of neuron parameters (weights and biases) that minimises this cost function. The optimization procedures and algorithms are already included in the aforementioned libraries ({\tt PyTorch} and/or Tensorflow).

\begin{figure*}
\centering
\includegraphics[width=16cm]{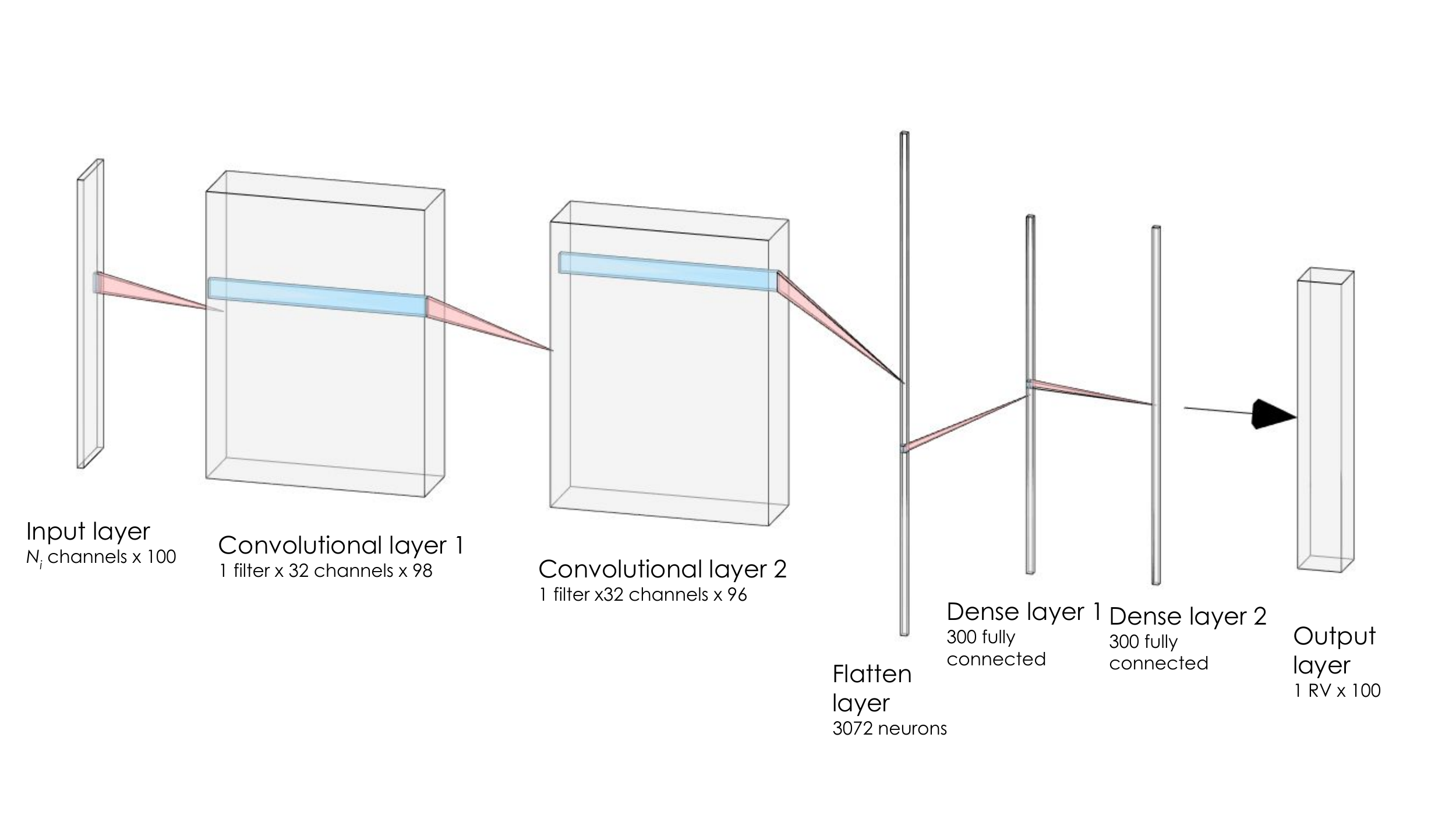}
\caption{Neural network architecture used throughout this study. The input layer consists of 100 measurements of $N_{\rm i}$ activity indicators, and the output layer (or label) are the corresponding 100 RVs. The first two hidden layers are convolutional layers, each with 32 kernels applied onto the three neighboring neurons. After the second convolutional layer, all the outputs for all the neurons are organised in a linear vector of neurons, and three subsequent fully connected layers are used to transform this vector into the final RV prediction. The response functions of all the neurons are the so-called Rectified Linear Unit functions (or ReLu), which clearly outperform other non-linear functions in terms of convergence speed and smaller computational time. As discussed in the text, small modifications of this architecture do not significantly change the obtained results.} \label{fig:NNarch}
\end{figure*}

For training, we use 90\% of the synthetic sets, and 10\% of the sets are reserved for testing and validation. On average, we found that these architectures would reach reasonable convergence (as measured by non-significant improvement of the cost function after 5 iterations) with about 50 training epochs. As an important check, we verified that the cost function applied to the test sets would always be equal (or slightly worse) than on the training sample, thus ensuring that we were not incurring in over-fitting. We show a visual representation of the NN architecture as described here in Fig.\,\ref{fig:NNarch}, as well as some computing-time benchmark experiments in Table\,\ref{table:batch}.

\section{Results} \label{sec:results}

In this section we perform a series of experiments to benchmark the inversion performance of different activity indicators. Each experiment consists of taking a time sampling (uniform, random, gap) and using a combination of activity indicators as the input for the NN. Then, the RRE of the RVs is computed for the full ensemble of outputs and used as a figure of merit of the procedure. We first do it with simulated datasets without uncertainties in the input nor the output RVs (section \ref{sec:NNsyn}), and then we replicate all the experiments for the uniform time sampling adding realistic noise (using the error value from Table\,\ref{table:timeseries}) to the inputs to quantify the impact of uncertainties in the training set and the overall performances (section \ref{sec:NNjitter}).

From the first experiments, it became clear that the most computationally intensive part of this study was the simulation of the synthetic data sets and not the training of the NN. With our setup, simulating one data set consisting of RV, FWHM, BIS, CON, $V$, $B$ and $I$ takes about one second per CPU. Assuming 8 cores, it takes about 1.5 days to complete $10^6$ simulations. These simulations can be easily parallelised, the total computing time being proportional to the inverse of the number of CPUs employed. For illustrative purposes, Table \ref{table:batch} shows the computer time needed to perform one training iteration, showing that these are relatively small (the 50 epoch values that we use to train each NN can be reached within a few minutes). The table illustrates the advantage of using GPU enabled computers and choosing a suitable size of a parameter called batch-size, which is the number of samples used in a single optimization step within one training epoch. This is a key parameter to take full advantage of the parallel computing power of GPUs as abundantly discussed in NN literature. A large batch size uses many samples to adjusts all the parameters of an NN (sends a large number of samples to the GPU to perform the back propagation adjustment of the NN), but may smooth out subtle effects contained in the simulations thus making it more difficult to converge to an optimal solution. We found that an optimal batch size of 512 samples per training step was sufficient to reproduce the convergence performance of smaller batch sizes while keeping a the computing time quite low.

\begin{table}
\caption{Numerical performance tests as a function of batch size.
\label{table:batch}}
\centering           
\begin{tabular}{c r r | c c}     
\hline\hline \noalign{\smallskip} 
Batch &  Computer &  Number & GPU & CPU \\ 
size & time per & of & usage & usage \\ 
& epoch [sec] & batches & [\%] & [\%] \\ 
\hline \noalign{\smallskip}
\multicolumn{5}{c}{CPU only} \\
\hline \noalign{\smallskip}
  32 & 120  & 28125 & 0 & 90\\
  64 &  82 & 14063 & 0 & 90\\
 128 &  81 &  7032 & 0 & 90\\
 256 &  66 &  3516 & 0 & 92\\
 512 &  60 &  1758 & 0 & 93\\
\hline \noalign{\smallskip}
\multicolumn{5}{c}{CPU + GPU} \\
\hline \noalign{\smallskip}
  32 & 92  & 28125 & 2  & 14 \\
  64 & 55 & 14063 & 5  & 17 \\
 128 & 28 & 7032  & 5  & 17 \\
 256 & 13 & 3516  & 6  & 10 \\
 512 &  8 & 1758 &  5  & 10 \\
1024 &  6 &  879 &  5  & 10 \\
2048 &  5 &  440 &  5  & 10 \\
\noalign{\smallskip}
\hline   
\end{tabular}
\tablefoot{For this experiment in particular, we generated one million synthetic data sets following the time-sampling of \epseri (66 observations). For each experiment we trained a convolutional neural network (three convolutional layers, followed by two fully connected ones). The input samples consisted of three spectroscopic indices (FWHM, BIS, and CON).}
\end{table}

\subsection{Neural Network training without observational errors} \label{sec:NNsyn}

\begin{figure}
\centering
\includegraphics[width=9cm]{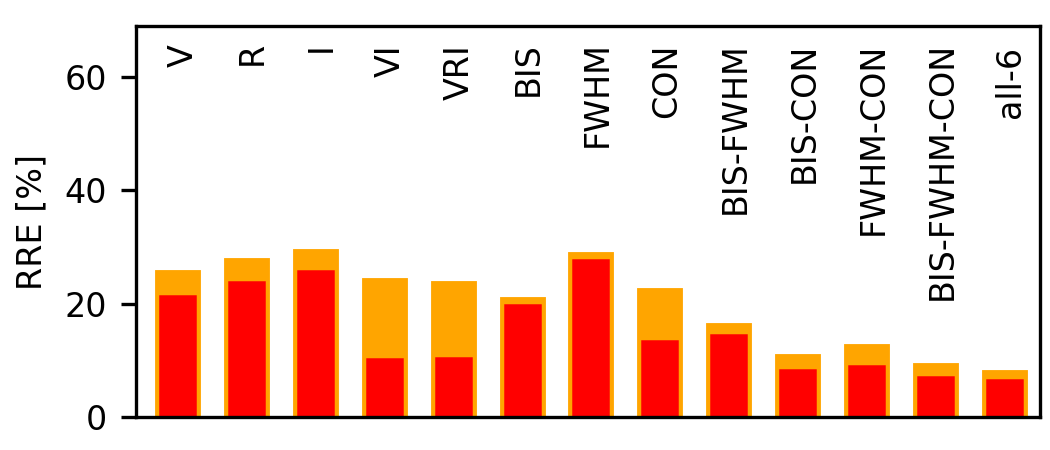}
\includegraphics[width=9cm]{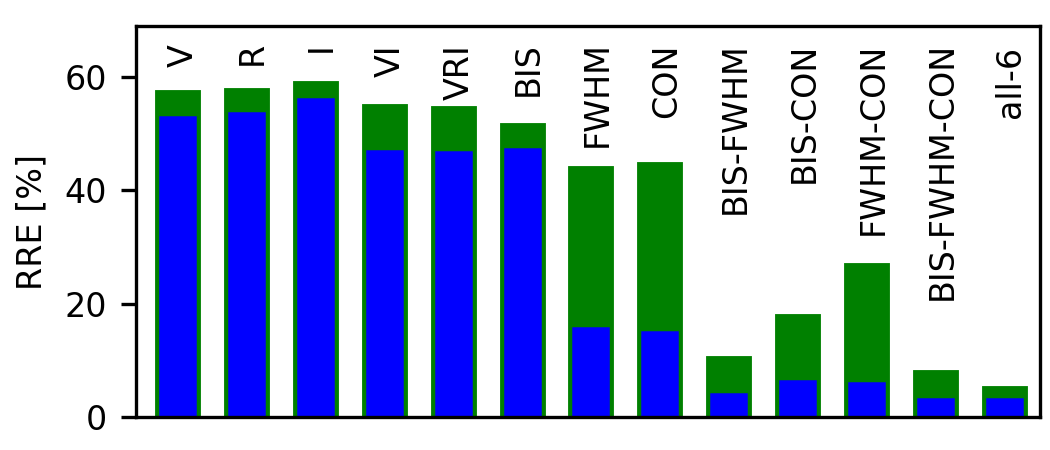}
\caption{Relative residual error (RRE) after the application of the trained NNs on the 540\,000 and 900\,000 {\tt starsim} models in percentage for \epseri (top) and \aumic (bottom), respectively. We mark the different activity-index combinations used for the neural network training. The average RV reduction for the different time samplings \emph{average} is given for \epseri in red and for \aumic in blue. The yellow and green bars indicate the value, if the relative additional error from Table\,\ref{table:timeseries} was added to the input data.} \label{fig:ResultsTable}
\end{figure}

\begin{figure}
\centering
\includegraphics[width=9cm]{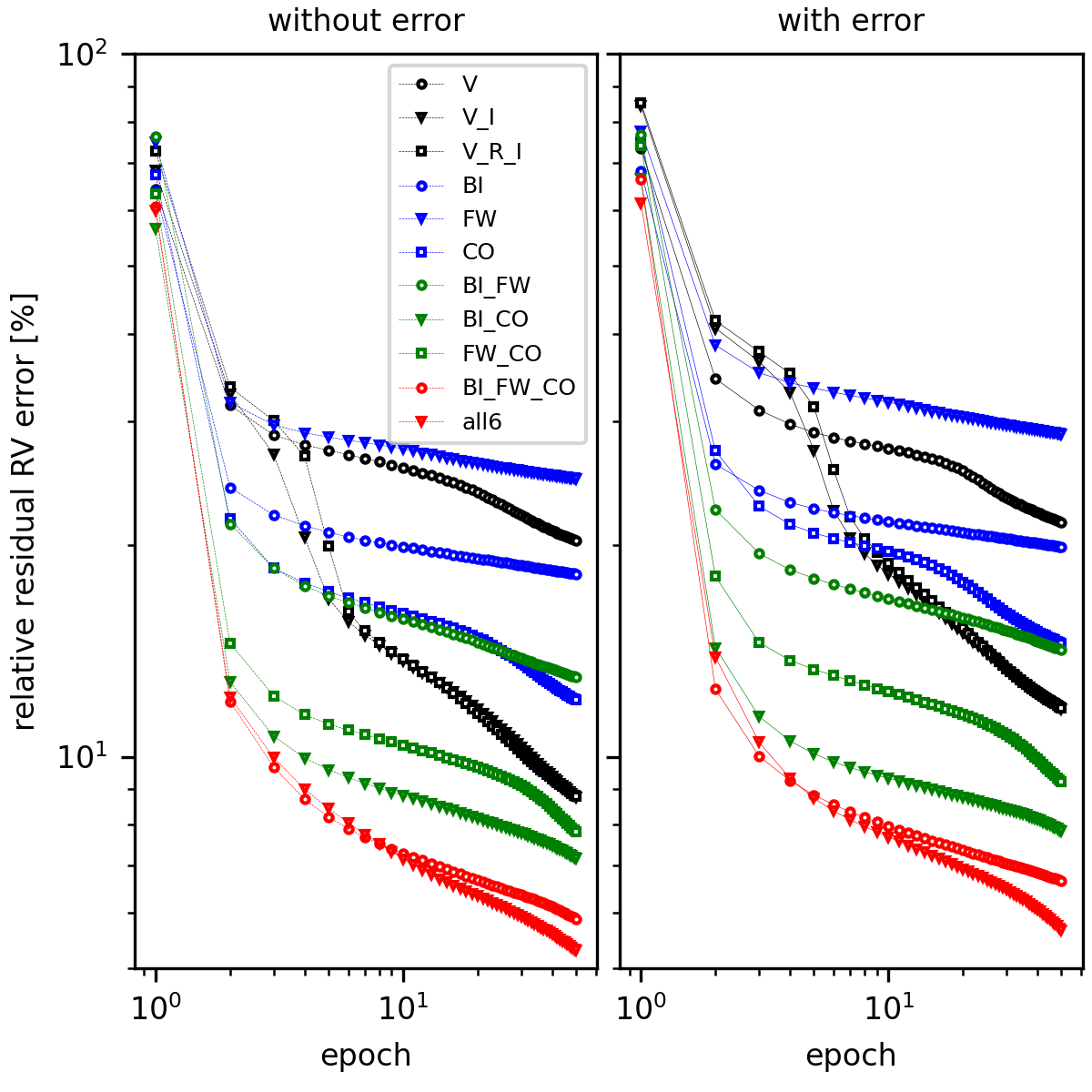}
\caption{Evolution of the relative residual error (RRE) of the 60\,000 synthetic \epseri test set RVs for the 50 epochs for the neural networks using all different combinations of input data, i.e. activity index time series as indicated, the uniform time sampling, and no additional error on the training input sets. The curves are shown for the test sets without additional uncertainties as calculated by {\tt starsim} (left panel, dashed lines), and for the case, where we included the error value from Table\,\ref{table:timeseries} for each activity index into the test sets (right panel, full drawn lines).} \label{fig:EELoss}
\end{figure}

\begin{figure}
\centering
\includegraphics[width=9cm]{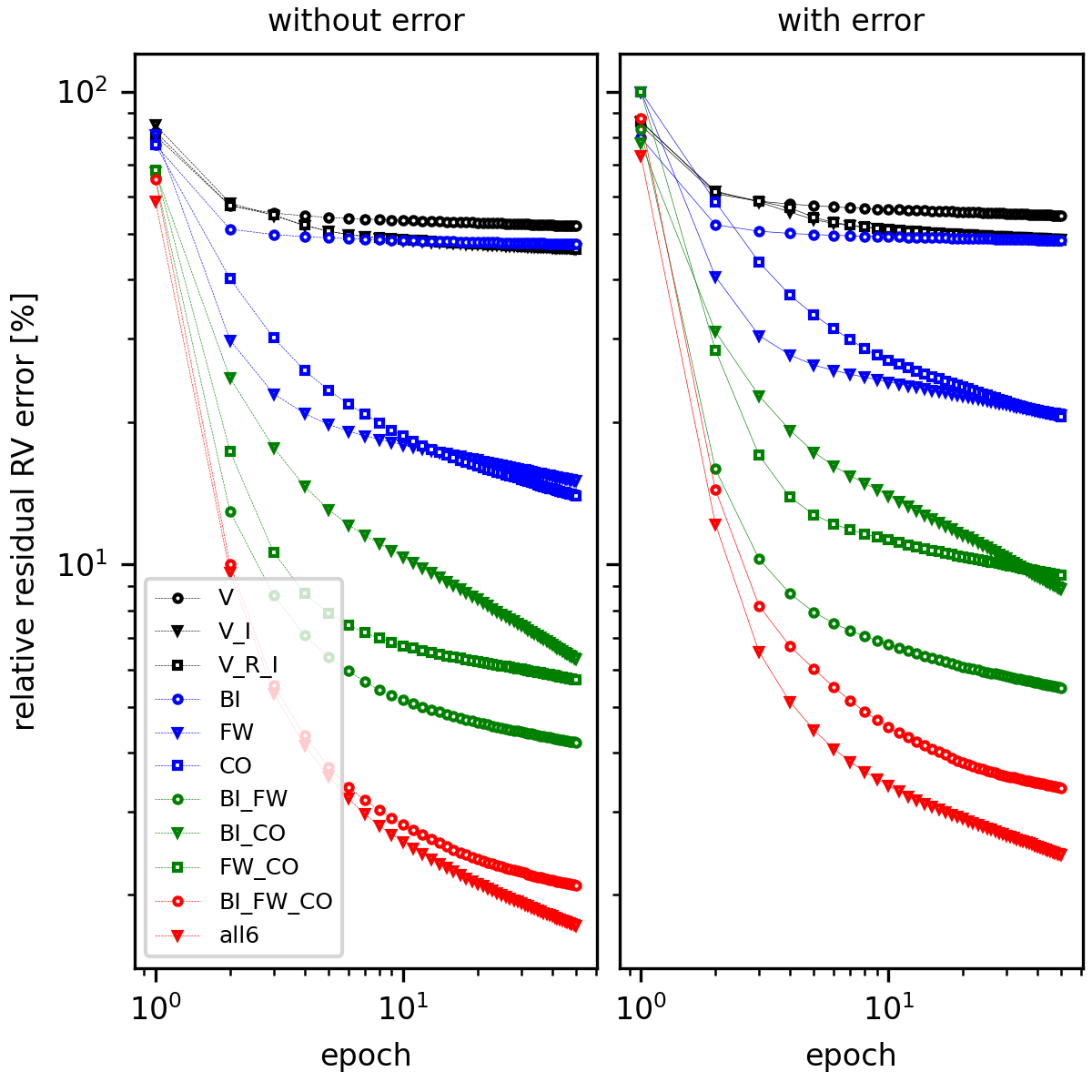}
\caption{Evolution of the relative residual error (RRE) of the 100\,000 synthetic \aumic test set RVs for the 50 epochs for the neural networks using all different combinations of input data. See Fig.\,\ref{fig:EELoss} for more details.} \label{fig:AUloss}
\end{figure}

Tables \ref{table:resultsEpsEri} and \ref{table:resultsaumic} show the RRE for each exercise, i.e. the relative residual RV error. The combinations of activity indices used in each exercise are shown in the left column. We show further the values averaged over all time samplings (\emph{average}), and the results obtained for the \emph{uniform} time sampling when we inject realistic noise (see Table\,\ref{table:timeseries}) to the input indices (see Sect.\,\ref{sec:NNjitter}). Figure\,\ref{fig:ResultsTable} visually shows the same results. Additionally, the RRE of the test sets as a function of the number of epochs is shown in Fig.~\ref{fig:EELoss} and Fig.~\,\ref{fig:AUloss} for both stars. There, we used the NNs trained with the uniform time sampling and without additional errors, and applied it to the test sets without (left panels) and with (right panels) the additional errors for the test input sets, respectively.

\begin{table*}
\caption{Relative residual error (RRE) of the RVs in \% after 50 training runs for the neural networks using 540\,000 {\tt starsim} models for different time samplings and for different kinds and numbers of activity indices as input for \epseri.} \label{table:resultsEpsEri}
\centering                         
\begin{tabular}{l r r r r r r r}     
\hline\hline \noalign{\smallskip} 
 Input index& \emph{uniform} & \emph{random1} & \emph{gap1} & \emph{random2} & \emph{gap2} & \emph{average} & with error \\
\noalign{\smallskip}  \hline \noalign{\smallskip} 
$V$ & 19.9 & 20.7 & 18.7 & 21.3 & 20.3 & 20.2$\pm$0.4 & 23.8 \\
$R$ & 22.4 & 22.9 & 21.4 & 23.7 & 23.0 & 22.7$\pm$0.3 & 26.0 \\
$I$ & 23.9 & 24.7 & 24.2 & 25.5 & 24.7 & 24.6$\pm$0.2 & 27.5 \\
$V$-$I$ & 8.6 & 9.5 & 8.6 & 9.9 & 8.8 & 9.1$\pm$0.2 & 22.4 \\
$V$-$R$-$I$ & 8.7 & 9.6 & 8.5 & 10.0 & 8.8 & 9.1$\pm$0.2 & 22.0 \\
BIS & 17.0 & 19.3 & 18.8 & 19.7 & 18.2 & 18.6$\pm$0.4 & 19.1 \\
FWHM & 23.1 & 29.1 & 27.0 & 28.5 & 24.9 & 26.5$\pm$1.0 & 27.1 \\
CON & 12.6 & 12.4 & 10.4 & 13.1 & 12.0 & 12.1$\pm$0.4 & 20.6 \\
BIS-FWHM & 11.9 & 13.9 & 13.4 & 14.0 & 13.0 & 13.2$\pm$0.3 & 14.5 \\
BIS-CON & 6.5 & 7.8 & 6.1 & 7.8 & 7.2 & 7.1$\pm$0.3 & 9.1 \\
FWHM-CON & 7.5 & 7.9 & 6.6 & 8.7 & 7.8 & 7.7$\pm$0.3 & 10.8 \\
BIS-FWHM-CON & 5.1 & 5.8 & 5.5 & 6.5 & 5.9 & 5.8$\pm$0.2 & 7.5 \\
all 6 & 4.7 & 5.7 & 4.9 & 5.6 & 5.3 & 5.2$\pm$0.2 & 6.1 \\
\noalign{\smallskip}
time-BIS & 17.0 & 19.2 & 18.9 & 19.7 & 18.2 & 18.6$\pm$0.4 & 19.1 \\
\noalign{\smallskip} \hline                
\end{tabular}
\tablefoot{Column \emph{average} shows the average of all the values of the other time samplings and their standard deviation, \emph{with error} shows the RV RRE of the \emph{uniform} time sampling, where the additional relative residual errors (see Table\,\ref{table:timeseries}) are added to each index of the NN input data.}
\end{table*}

\begin{table*}
\caption{Relative residual error (RRE) of the RVs in \% after 50 training runs for the neural networks using 900\,000 {\tt starsim} models for different time samplings and for different kinds and numbers of activity indices as input for \aumic.} \label{table:resultsaumic}
\centering                         
\begin{tabular}{l r r r r r r r}     
\hline\hline \noalign{\smallskip} 
 & \emph{uniform} & \emph{random1} & \emph{gap1} & \emph{random2} & \emph{gap2} & \emph{average} & with error \\
\noalign{\smallskip}  \hline \noalign{\smallskip} 
$V$ & 55.8 & 53.5 & 52.2 & 54.1 & 51.9 & 53.5$\pm$0.6 & 57.4 \\
$R$ & 56.2 & 54.4 & 53.0 & 54.9 & 52.7 & 54.2$\pm$0.6 & 57.8 \\
$I$ & 57.7 & 57.2 & 55.9 & 57.5 & 55.6 & 56.8$\pm$0.4 & 59.0 \\
$V$-$I$ & 48.9 & 47.7 & 46.6 & 48.1 & 46.6 & 47.6$\pm$0.4 & 55.0 \\
$V$-$R$-$I$ & 48.9 & 47.5 & 46.3 & 48.1 & 46.5 & 47.4$\pm$0.4 & 54.6 \\
BIS & 48.2 & 47.9 & 47.8 & 48.0 & 47.6 & 47.9$\pm$0.1 & 51.7 \\
FWHM & 19.2 & 16.0 & 14.9 & 16.6 & 15.0 & 16.3$\pm$0.7 & 44.0 \\
CON & 21.7 & 14.3 & 11.8 & 16.4 & 14.0 & 15.6$\pm$1.5 & 44.7 \\
BIS-FWHM & 4.7 & 4.6 & 4.4 & 4.7 & 4.2 & 4.5$\pm$0.1 & 10.4 \\
BIS-CON & 7.4 & 7.3 & 6.2 & 7.6 & 6.3 & 7.0$\pm$0.3 & 17.8 \\
FWHM-CON & 7.9 & 6.5 & 5.6 & 7.1 & 5.7 & 6.6$\pm$0.4 & 26.8 \\
BIS-FWHM-CON & 2.4 &s 2.5 & 2.3 & 2.5 & 2.1 & 2.4$\pm$0.1 & 8.0 \\
all 6 & 2.0 & 2.0 & 1.9 & 2.1 & 1.7 & 1.9$\pm$0.1 & 5.1 \\
\noalign{\smallskip} \hline             
\end{tabular}
\tablefoot{See notes on Table\,\ref{table:resultsEpsEri} for further details.}
\end{table*}

\begin{figure*}
\centering
\includegraphics[width=9cm]{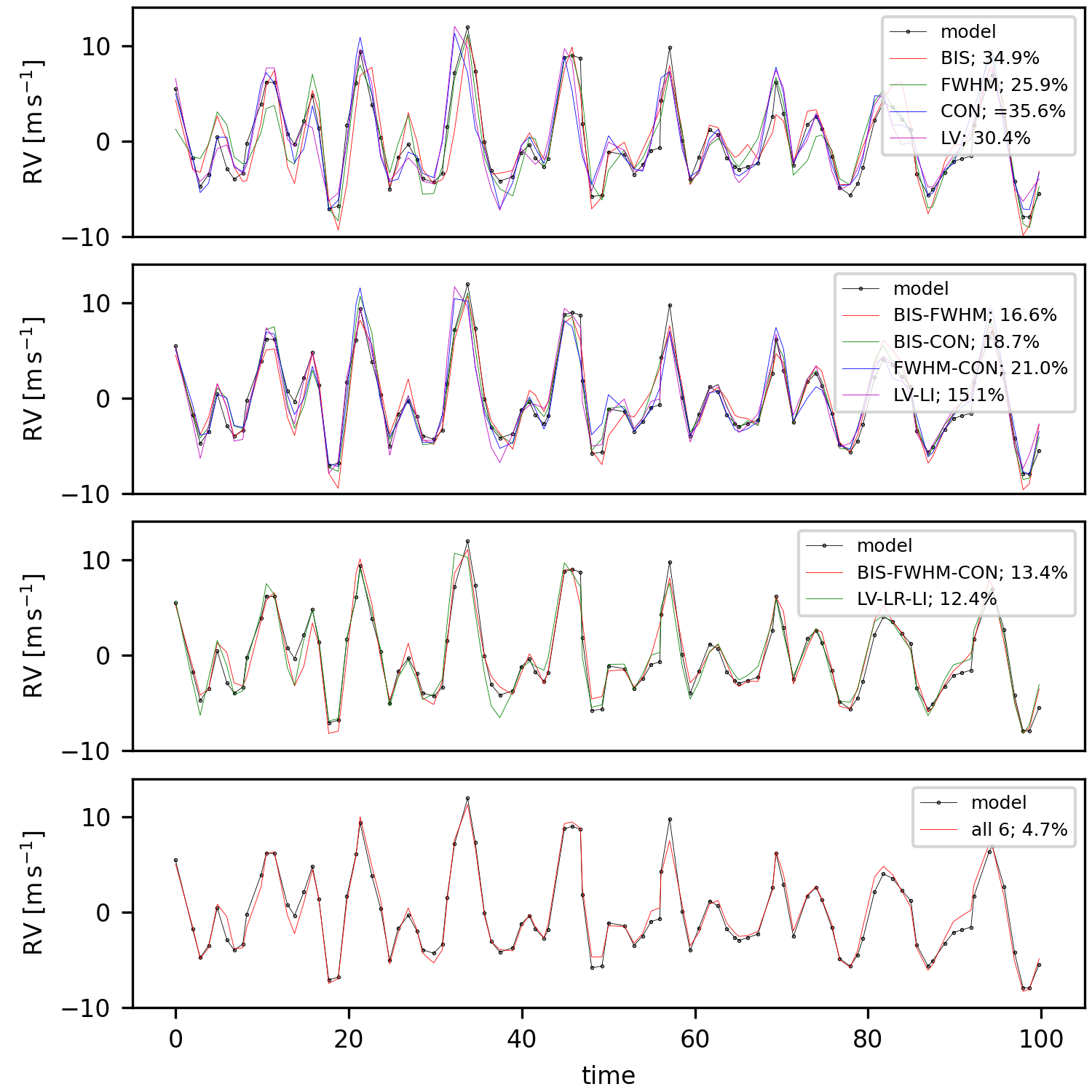}
\includegraphics[width=9cm]{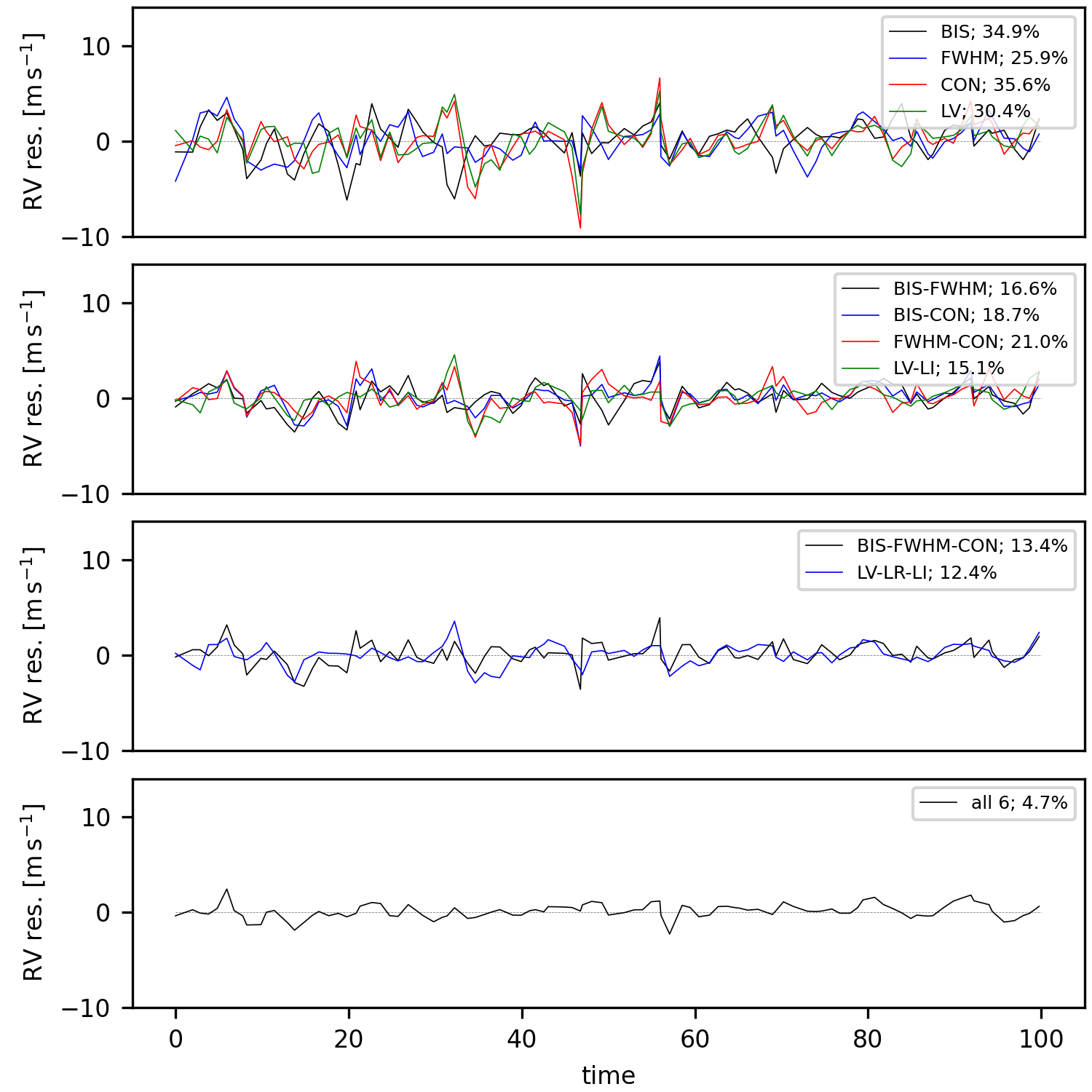}
\caption{Illustration of the quality of the RV correction as a function of the input indices used (random synthetic set chosen) for the \epseri data an the \emph{uniform} time-sampling (black lines). The output predictions of of the neural networks with all the combinations of input data, i.e. index time series are also illustrated with different colours. The outcome of using only one index is shown in the top panel, further down two indices, then combinations of three indices (photometry and CCF), and then all six index time-series data. RVs are shown as modeled by {\tt starsim}. The legends contain RV RRE of this particular set for each combination of index inputs. The residuals (differences between RV predictions and the simulated RV) are shown in the right panels.} \label{fig:randomSet}
\end{figure*}

As a further illustration, we show one random synthetic RV test set from the \emph{uniform} time sampling and the output of the NNs with all the combinations of input data, i.e. index time series, in the left panels of Fig.\,\ref{fig:randomSet}. From the top, we see in the first panel the usage of only one index, then two indices, then the three indices from photometry and from the CCF, and then all six index time-series data. The model itself is shown by the black dots and line. In the right panels, the difference between the RV model and its corresponding prediction by the NN is shown.

In general, we find that even using a single index, there is always a reduction in RRE. But even for data without uncertainties, not all the activity induced RV variability can be removed from the observations. Let us note that this \emph{inversion error} marks the floor of what is possible given the information in the corresponding index (these are intrinsic model degeneracies) and our NN implementation. Additional indices at the same epochs (additional spectroscopic observables) and-or additional non-simultaneous measurements (e.g. contemporaneous photometric observations) could improve these floor values, but one should expect worse performances when applied to real/noisy observations.

The most important take home messages from these experiments can be summarised as follows.

\begin{enumerate}

\item {\bf The proposed neural network architecture is sufficient} to produce high quality corrections to RV data. The NNs converge after only a few training epochs irrespective of the fine details of the NN algorithm (response function, training strategy or other technical details). The convergence and non-overfitting can be easily verified by checking that the RRE of the test sample (which is the merit function also used during training) is similar or slightly larger than the RRE of the training sample, which is always the case in all the experiments.

\item {\bf Line-profile indices contain enough information} to successfully train the algorithms, and produce a significant correction of the activity signals in the RV time-series. Despite the existence of correlations between RV and indices has been demonstrated in the past, this is the first time (to our knowledge) where it is shown that there is enough information in them to reproduce the RV activity signal.

\item {\bf The best performing single index was found to be CON, rather than BIS}, which would have been expected according to \citet{2020A&A...636A..36L}. This indicates that data/model driven approaches rather than guesswork or theoretical expectations are preferable in future activity/RV noise investigations. It is possible that other indices provide even better performances than those discussed here.

\item {\bf Combining multiple indices is the most effective way} to achieve a higher RV reduction, given that these measurements can be obtained from the same spectra used to obtain the RV measurement. The actual reduction, and the most relevant index is star dependent (as has been suspected for some time now) and, in general, one should aim at using a combination of all possible indices that show variability consistent with stellar activity. In particular, using all the available CCF indices, the simulations indicate that we can reach a RRE of 5.1, and 2.4\,\% for \epseri and AU\,Mic, respectively. This is better than half the value as for any single index for $\varepsilon$\,Eri, but is remarkably lower (eight times) for \aumic. 

\item {\bf Simultaneous photometric time-series seem to contain the least information of the tested indices}, worsening for the redder filters. Even when combined, they only match the performance of a single CCF index as input. Especially for AU\,Mic, the obtained RREs are only at the 55\,\% level, which is a rather poor performance. Using all the available filters, we can reach RREs of 9 and 49\,\% for \epseri and AU\,Mic, respectively. This is likely caused by intrinsic degeneracies when mapping photometric signals into putative spots (and corresponding RV signal), as often happens in light curve-only inversion cases. For example, a small equatorial spot can produce the same photometric signal than a small polar one, but its RV counterpart can be rather different. In any case, adding photometry always helps, and the putative use of photometric time samplings denser than the RV sampling (as often happens when using space-based photometry) remains to be investigated.

\item {\bf A larger number of data-points seem to improve overall performance}. Probably due to the smaller number of data points (37 versus 66) and the later spectral type, the RRE of the RVs across all input data combinations is more than double in the \aumic data compared to the \epseri sets, and this is despite the larger training data sample (540\,000 vs. 900\,000).

\item {\bf Injecting noise in the test sets does not improve performance} when applied to noisless sets. That is, when trained with synthetic data \emph{without errors}, the performance on test data \emph{with errors} we observed that the RV RRE is not as good, and comparable to training \emph{without errors} (as discussed in the next section). This is illustrated in Figs.\,\ref{fig:EELoss} and \ref{fig:AUloss}. 

\item {\bf The precise time sampling does not have a strong influence} on the outcome, even with seasonal gaps. We remark, however, that all the test samplings used here cover the rotation period of both stars quite well (several cycles, more than one point per cycle). More tests would be needed to re-analyse old survey data, which typically spans several years and with observations clustered in seasons of a few months each year.

\item {\bf The time-tags of the observations were NOT used} in the experiments presented here (Tables\,\ref{table:resultsEpsEri} \& \ref{table:resultsaumic}). Experiments ran with time-tags and all combinations of indices converged much more slowly than those without them. This is something we did not expect that requires further investigation. The outcome of a test including the time tags as inputs together with the BIS index for \epseri is shown in Table\,\ref{table:resultsEpsEri}.

\end{enumerate}

\subsection{Neural Network training with realistic observational errors} \label{sec:NNjitter}

Since noise is an intrinsic characteristic of real data that shall also be learned, we considered whether we should \emph{train} our NN with noisy synthetic observations. In the following list we summarise the conclusions drawn by running test with noisy training sets and presented in Tables\,\ref{table:resultsEpsEri} and \ref{table:resultsaumic}, and Figs.\,\ref{fig:ResultsTable}:

\begin{enumerate}

\item {\bf The performance of CCF indices} becomes quite affected when using noisy input indices (and their combinations) when their RRE exceeds 50\% (FWHM and CON for AU\,Mic). Uncertainties in activity indices are not as well understood as for RV measurements, so this calls for a quick examination of possible periodic signals in all the indices to be used before deciding whether an index shall be used in the training and later de-trending of real time series. A real case example is discussed later in the section devoted to \aumic real observations (Sec.\,\ref{sec:NNaumicn}), where we found that using the CON index (which does not show the same periodicities as the other indices) makes things worse.

\item {\bf Along the same line, uncertainties in the photometry at few \% level (which are quite easy to obtain, even from the ground), do not affect the inversion power of the photometry as an input}. To illustrate this in the tests presented here, the precision of the photometric time series was arbitrarily chosen to be at 10\% level, which is rather high (see Table\,\ref{table:timeseries}). Despite this high value, this does not influence the outcome very much confirming that the less inversion power of the photometry comes from intrinsic physical degeneracies. 

\item {\bf The RRE of an activity index should be 30--40\% smaller than the contribution to the RV RRE caused by stellar activity variability}. If larger (FWHM and CON for \aumic), the introduction of those errors degrades performances very much (RV RREs form 16 to 44\,\%, and 16 to 45\,\% in these example cases).

\end{enumerate}

\subsection{Application to observational data}  \label{sec:NNrealTS}

\subsubsection{Results on $\varepsilon$\,Eridani}

Before jumping into generating a final set of synthetic training sets, we perform an exact inversion using the {\tt starsim} code in order to verify that the priors set in our simulations matched what would be obtained if attempting a more classic full stellar fitting approach. This process is explained in detail in Appendix\,\ref{sec:inversion}. Most importantly, the {\tt starsim} inversion resulted in differences in the stellar parameters compared to the literature ones, especially the stellar radius. With this new parameters, we then produce 100\,000 {\tt starsim} simulations of RV, BIS, FWHM, and CON time series, each using the same original time-sampling consisting in 66 observations. For the spot map, we use the random procedure as used in previous sections to avoid producing simulations too specific to the data. To train the NN, we add uncertainties and errors to the inputs and outputs as described in previous sections since in the case for the real observational data the NN performance is significantly enhanced.

We then train our NNs again on 50 training epochs achieving an RV RRE down to 10\% on our 100\,000 training sets, which is slightly larger but comparable to the experiments (7.5\%) in the previous section.

\begin{figure}
\centering
\includegraphics[width=9cm]{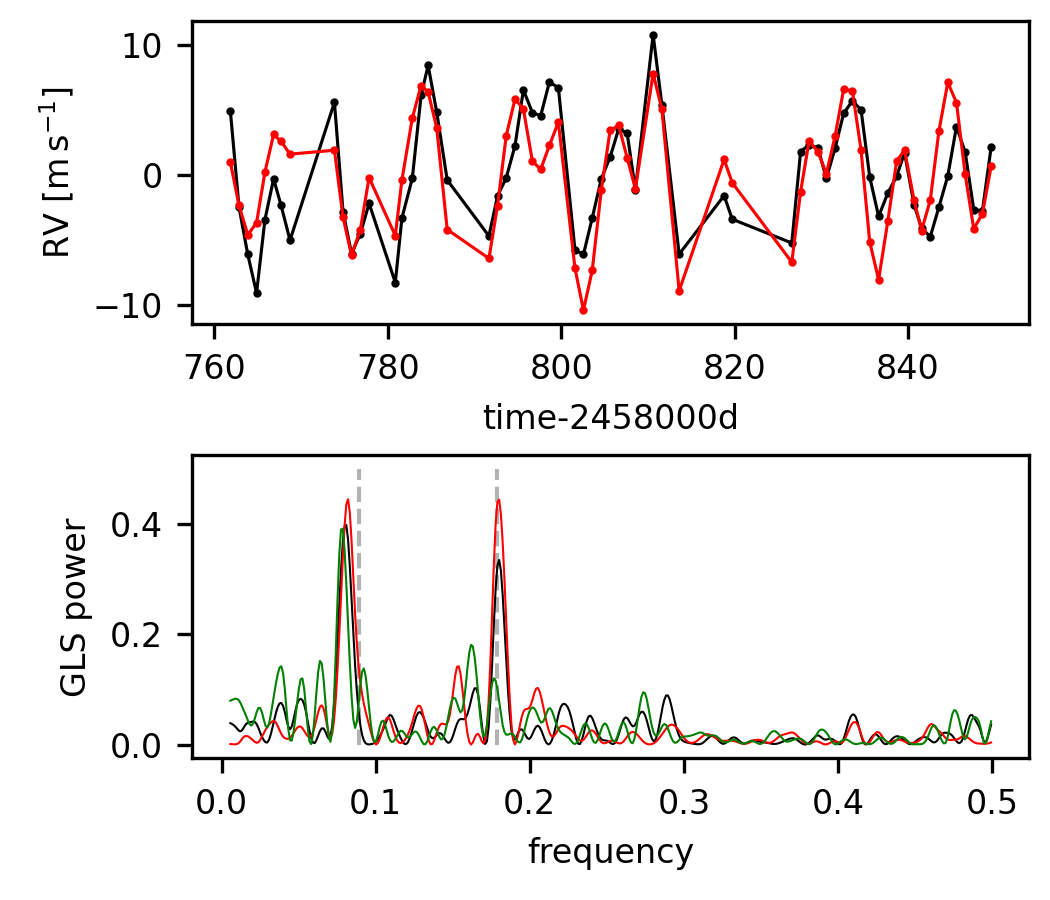}
\caption{RVs (top panel) and periodogram (bottom panel) of observational \epseri data in black. In red, we show the prediction produced by our neural network using BIS, FWHM, and CON indices as input. In green we show the periodogram of the \textit{cleaned} data, that is the difference between the observed and modeled RVs. The rotation period and its second harmonic are indicated as gray vertical dashed lines. The level of RMS reduction is about 45\%.} \label{fig:NNresultsRV}
\end{figure}

When applied to the real dataset, we obtain the correction shown in Fig.\,\ref{fig:NNresultsRV}. The black curve shows the observed RVs and the red curve shows the outputs of a trained NN with inputs of BIS, FWHM, and CON from the observations. The periodograms of the real and predicted RVs are also shown in the panel below. The corresponding RV RRE is \textit{45\%} (from 4.4\,m\,s$^{-1}$ to 2.0\,m\,s$^{-1}$) which is quite far from the nominal 10\% inferred from the training and test samples. We argue that at the level below few m\,s$^{-1}$, the physics of the activity induced effects are not yet properly included in the simulations and, possibly, some instrumental systematic errors are also affecting the result (e.g. uncertainty in the wavelength solution of each night of observation). More work on the models is needed to identify all possible sources of such mismatches. In any case, the reduction of the RMS below 2\,m\,s$^{-1}$ already places \epseri as a high quality target (similar ``jitter'' as the Sun) for exoplanet searches (enough sensitivity to detect close-in super-Earths).

\subsubsection{Results on AU\,Microscopii} \label{sec:NNaumicn}

We apply the same procedure to the case of AU\,Mic, but without performing the initial inversion with {\tt starsim}. Instead, we use the stellar parameters as introduced in Table\,\ref{tab:teststars}, the original time sampling containing 37 data points, and randomised spot maps as introduced in Sect.\,\ref{subsec:starsim}. 

\begin{figure}
\centering
\includegraphics[width=9cm]{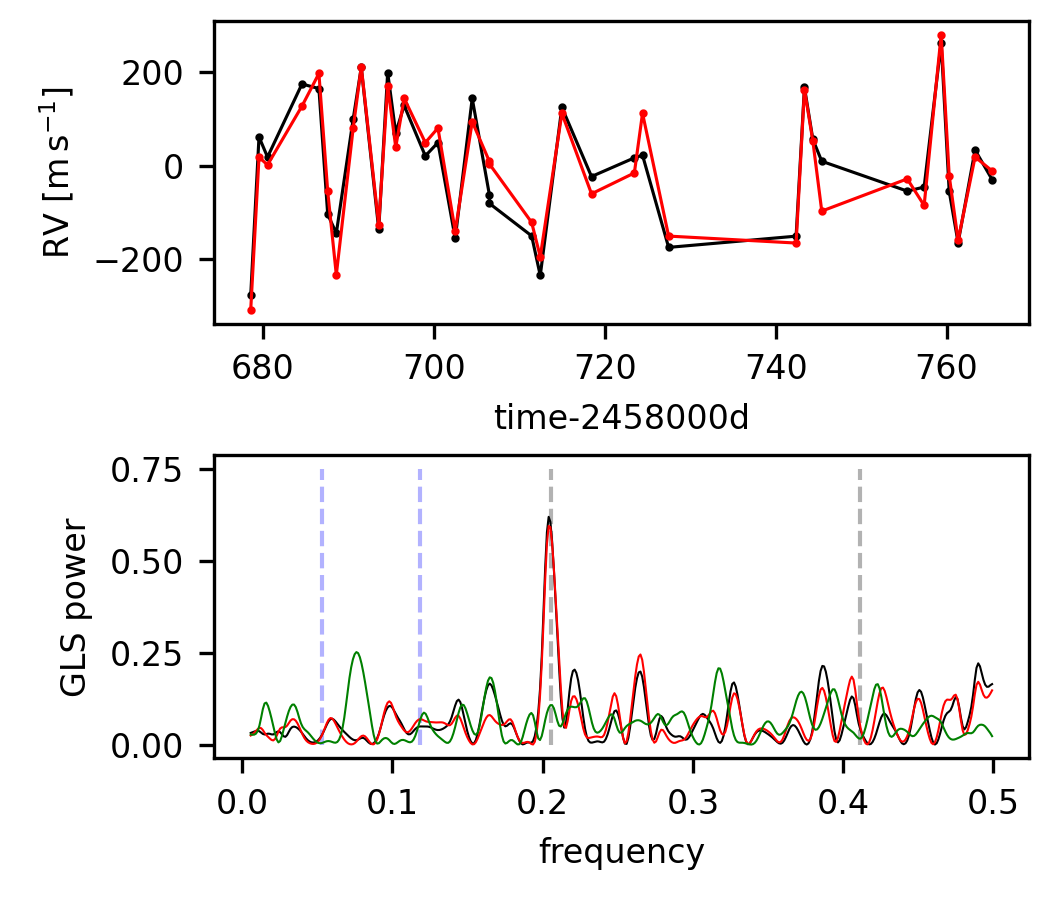} 
\caption{top panel: RVs from \aumic as observed (black) and as modeled by the trained neural network using BIS, and FWHM indices with their respective relative residual errors. bottom panel: periodogram of the observed RVs (black), the neural network output (red), and the difference (green). Black dashed lines indicate the rotation period and its second harmonic, blue dashed lines indicate the orbits of the two planetary companions of AU\,Mic. We reach down to RREs of 10\%.} \label{fig:aumicSampl}
\end{figure}

For this experiment, we create 400\,000 new {\tt starsim} simulations calculating RV, BIS, and FWHM, and we include again the observational errors (see Table\,\ref{table:timeseries}) into the time series since it has been shown that it results in lower RV RREs. We then train our NN for 50 training epochs as before and reach an RV RRE of 10\% for the test set. This is essentially the same value as in the previous test. We show the RV observations and the NN output in the top panel of Fig.\,\ref{fig:aumicSampl}. In the bottom panel, we show the periodograms of those two time series and add in green the periodogram of the difference.

As opposed to the expectations from all previous experiments, the inclusion of the CON index produced very poor performances when applied to the real data. After a close examination of the CON measurements and corresponding periodograms (see Fig.\,\ref{fig:timeseries2}) it is clear that they do not contain the same characteristic periodicities as the other indices, or the RV. We attribute this to the fact that CON measurements from CARMENES are affected by additional noise sources not included in the simulations, including instrumental and astrophysical effects. For example, CON measurements are very sensitive to the continuum determination (and corresponding line-depths of the lines). Therefore, the source of this systematic error might be instrumental (e.g. change in the instrumental setup such as using simultaneous calibration lamps, sky background contribution, change in gain in the detector), but it could also be associated to astrophysical phenomena causing changes in the spectral continuum. For these reasons, we only use the BIS and FWHM indices for the performance analysis on real observations of \aumic.

\begin{figure}
\centering
\includegraphics[width=9cm]{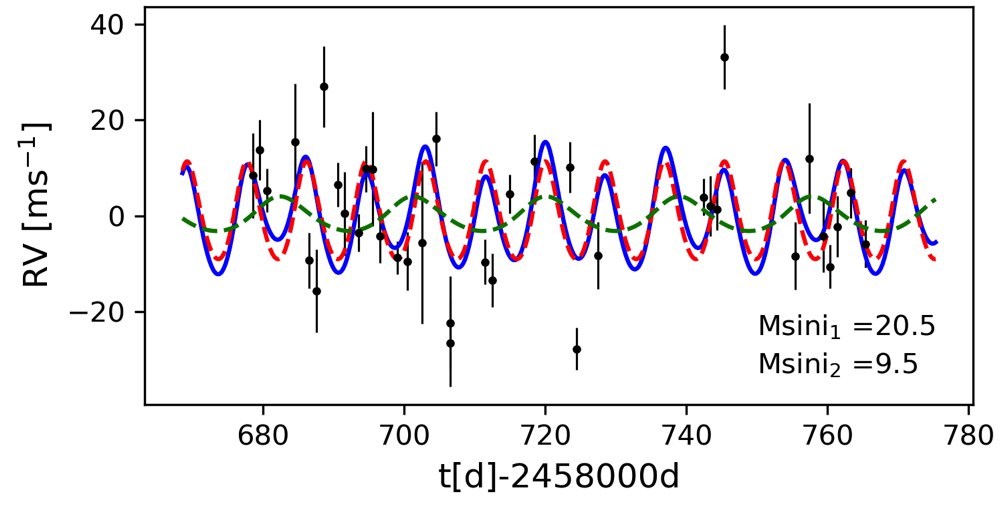} 
\caption{Residual RVs of the 37 \aumic observations after applying the neural network using BIS and FWHM indices (black dots). We further show the curves introduced to the RVs by the two detected planets of \aumic (red and green dashed lines) and their combination (blue curve).} \label{fig:AUresidual}
\end{figure}

The correction obtained on the RVs is rather good, reaching to a RV RRE down to 10\%, the same as for the {\tt starsim} training sets. The cleaned RVs (subtraction of the computed correction to the observed ones), do not contain any trace of the rotation period. Moreover, the resulting scatter is consistent with the presence of the Doppler signals of the planets (although there are not enough observations to confirm them with only these 37 epochs). As opposed to $\varepsilon$\,Eri, most of the activity induced variability is clearly related to the dark spot effect on a fast rotating star (peak to peak variability over 100\,m\,s$^{-1}$), which is one of the better understood RV noise phenomena. Although the relative improvement is rather good, we also note that the \textit{residual} RV variability is around 13\,m\,s$^{-1}$, which is substantially poorer than the one from $\varepsilon$\,Eri. This is is again consistent with the models lacking relevant physical realism at the few m\,s$^{-1}$ level. Using other de-trending techniques and evidence of periodic transits from photometric observations \citep{2021AJ....162..295C}, the possible amplitude of two low mass companions have been inferred at the level of $\sim$10\,m\,s$^{-1}$, which could explain part of this variability as well (Keplerian orbits have not been included in the training samples in any case). We show the residual RVs in Fig.\,\ref{fig:AUresidual}, and illustrate the possible contribution of the signals of the two published planetary companions to the residual RV noise (red and green dashed curves for the individual planets, and the blue curve for their combination).

\section{Conclusions} \label{sec:conclusions}

We developed a model-driven machine learning approach using convolutional deep neural networks to produce corrections to radial velocity time series contaminated by stellar activity. It is \textit{model driven} in the sense that synthetic data sets were used to train the algorithms. Stellar activity manifests itself in other observables than RVs and this additional information was already used previously to mitigate the influence of activity on the RVs. However, it was not entirely clear whether the information contained in the individual and combined observables was sufficient to invert the problem and produce a correction in a more general and independent way, not using the information contained in the RVs. This work applied to the test data sets demonstrates that the inversion is possible and that it is solvable, without requiring explicit expressions for all the relations of the indices with the RV variability.

We identified neural network architectures that are well suited to the task (a few convolutional layers followed by a few dense layers) and that they can be trained with quite modest computational resources. Actually, producing realistic simulations is the most computer intensive part of the processing. 

We assessed the inversion power of the classic activity indices used in exoplanet detection papers, and found that combinations of them are sufficient to provide corrections below the photon noise of state-of-the-art spectrometers when we consider the effects of spots combined with rotation and convective blueshift only ($\sim 10$\,cm\,s$^{-1}$, the goal of ESPRESSO). Photometric information obtained simultaneously with the observations did not have that much inversion power, but on its own can still produce significant reduction of RV noise. Non-simultaneous (space-based) multi-band photometry is likely to perform better but it will require additional work and, possibly, slightly different neural network architectural choices. 

We applied our technique to two real datasets achieving a significant reduction of the activity induced RV variability, but the results are not yet at the level of the photon noise of the observations. This is likely caused by astrophysical phenomena yet to be included in the simulations, and due to systematic (instrumental and algorithmic) errors in the measurements of both the activity indicators and the RVs.

As a future step, we plan to test the Machine Learning framework with more models that include additional activity phenomena. In particular, models will have to include more complex phenomenae such as non-uniform convective flows (granulation, supergranulation, meridional flows), magnetic distortion of spectral lines (Zeeman splitting), different sensitivities of some spectral lines to stellar activity, and the effect of bright and chromospherically active regions to apparent Doppler shifts (plages, faculae or others).

\begin{acknowledgements}
  CARMENES is an instrument at the Centro Astron\'omico Hispano en Andaluc\'ia (CAHA) at Calar Alto (Almer\'{\i}a, Spain), operated jointly by the Junta de Andaluc\'ia and the Instituto de Astrof\'isica de Andaluc\'ia (CSIC).
  
  CARMENES was funded by the Max-Planck-Gesellschaft (MPG), 
  the Consejo Superior de Investigaciones Cient\'{\i}ficas (CSIC),
  the Ministerio de Econom\'ia y Competitividad (MINECO) and the European Regional Development Fund (ERDF) through projects FICTS-2011-02, ICTS-2017-07-CAHA-4, and CAHA16-CE-3978, 
  and the members of the CARMENES Consortium 
  (Max-Planck-Institut f\"ur Astronomie,
  Instituto de Astrof\'{\i}sica de Andaluc\'{\i}a,
  Landessternwarte K\"onigstuhl,
  Institut de Ci\`encies de l'Espai,
  Institut f\"ur Astrophysik G\"ottingen,
  Universidad Complutense de Madrid,
  Th\"uringer Landessternwarte Tautenburg,
  Instituto de Astrof\'{\i}sica de Canarias,
  Hamburger Sternwarte,
  Centro de Astrobiolog\'{\i}a and
  Centro Astron\'omico Hispano-Alem\'an), 
  with additional contributions by the MINECO, 
  the Deutsche Forschungsgemeinschaft through the Major Research Instrumentation Programme and Research Unit FOR2544 ``Blue Planets around Red Stars'', 
  the Klaus Tschira Stiftung, 
  the states of Baden-W\"urttemberg and Niedersachsen, 
  and by the Junta de Andaluc\'{\i}a.

  We acknowledge financial support from the Agencia Estatal de Investigaci\'on of the Ministerio de Ciencia e Innovaci\'on MCIN/AEI/10.13039/501100011033 and the ERDF ``A way of making Europe'' through projects 
  PID2020-120375GB-I00,         % DMLEXO - ML grant - Barcelona  
  PID2019-109522GB-C5[1:4],	    % CAB+IAA+IAC+UCM
  PGC2018-098153-B-C33,         % ICE
  and the Centre of Excellence ``Severo Ochoa'' and ``Mar\'ia de Maeztu'' awards to the 
  Institut de Ciències de l'Espai (CEX2020-001058-M)
  Instituto de Astrof\'isica de Canarias (CEX2019-000920-S), Instituto de Astrof\'isica de Andaluc\'ia (SEV-2017-0709), and Centro de Astrobiolog\'ia (MDM-2017-0737);
  the Generalitat de Catalunya/CERCA programme;
  and the DFG through priority program SPP 1992 ``Exploring the Diversity of Extrasolar Planets'' (JE 701/5-1). % Jeffers

Based on observations collected at the European Organisation for Astronomical Research in the Southern Hemisphere under ESO programme 0104.C-0863(A); also called Red Dots. We also thank the Red Dots members Carole Haswell, Mario Damasso, Fabio del Sordo and Stefan Dreizler for their participation in the preparation of the Red Dots ESO observations with HARPS, their comments and insightful discussions.

\end{acknowledgements}

\bibliographystyle{aa}
\bibliography{bibtex.bib} 

\begin{appendix}

\section{{\tt starsim} data inversion for $\varepsilon$\,Eridani} \label{sec:inversion}

\begin{figure}
\centering
\includegraphics[width=10cm]{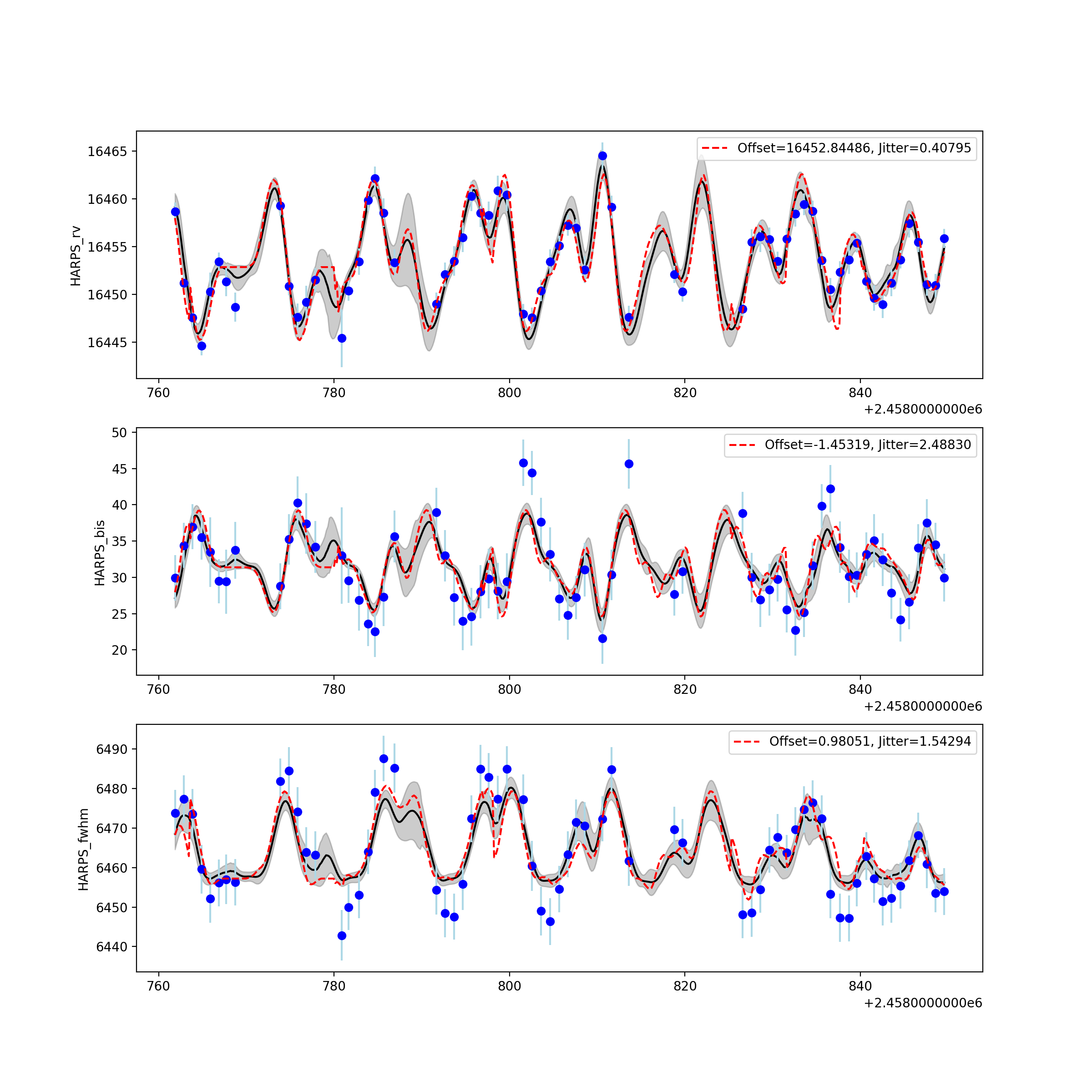}
\caption{Results of the {\tt starsim} inversions for \epseri using RV (top panel), FWHM (middle panel), and BIS (bottom panel). Blue dots indicate the measurements as extracted from {\tt raccoon} from the observed spectra. The black curve shows the average fit and the gray area the uncertainty of the different solutions. The dashed red curve shows the solution with the largest likelihood. On the top right, the offset for each time series is indicated as well as its additional jitter both showing the shortcomings of the modeling of time-series data with {\tt starsim}.} \label{fig:inversion_fit}
\end{figure}

\begin{figure}
\centering
\includegraphics[width=9cm]{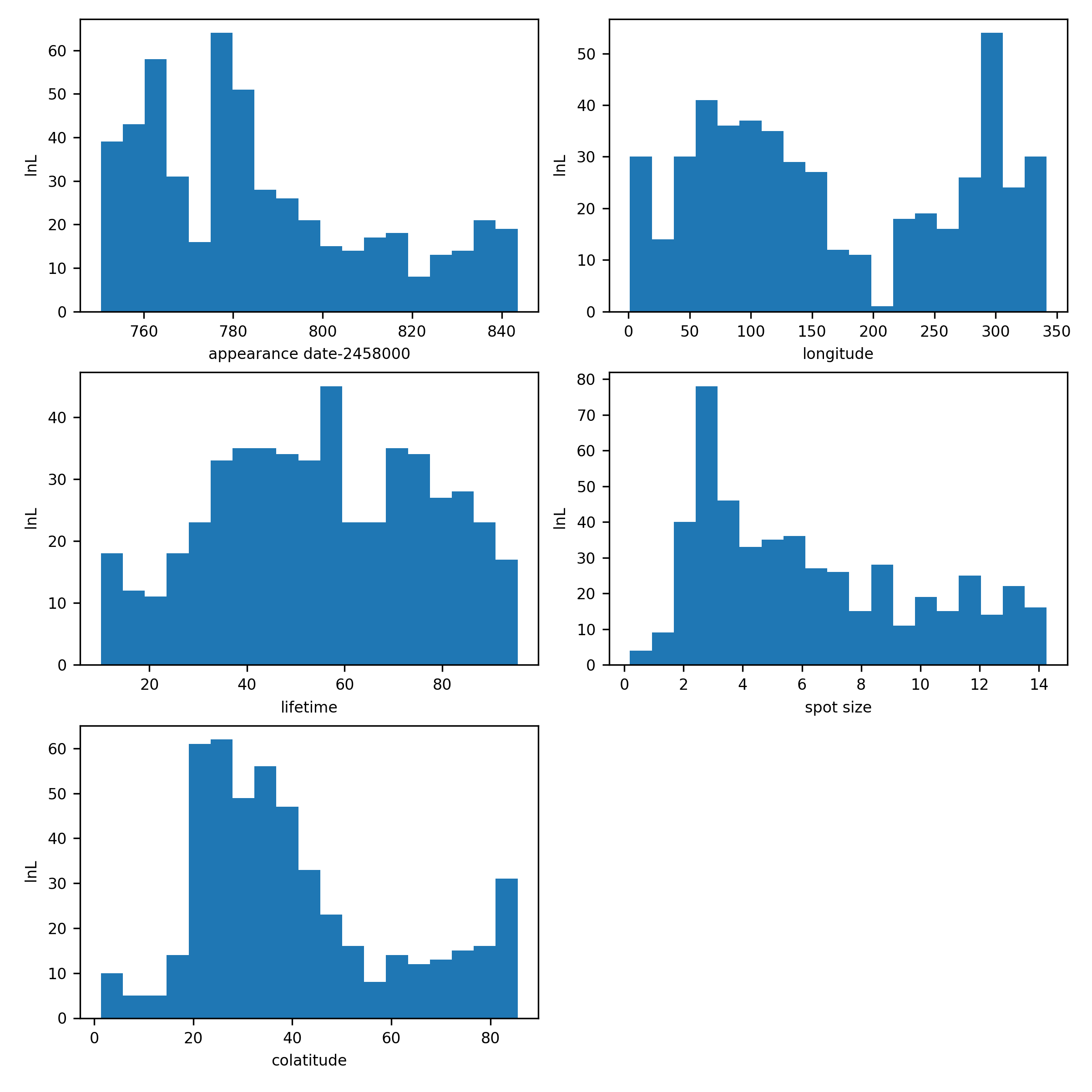} 
\caption{Histogram of the posterior distribution of the five parameters of the spot maps containing 10 spots of the {\tt starsim} inversion applied to observational \epseri data RV, BIS, and FWHM.} \label{fig:spotdis}
\end{figure}

With the current version of the {\tt starsim} code we calculate the inversion of the algorithm, i.e. using the observational time-series data of \epseri to fit for stellar parameters and an average distribution of stellar spots. Differently from our initial approach, we use a fixed number of 10 different dark spots up to 15\,$^{\circ}$ in radius, appearing at any point of the calculated time interval randomly and stellar parameters floating around the measurements as given in Table\,\ref{tab:teststars}. 
As a result, we achieve a best fit on observational RV, BIS, and FWHM time series as shown in Fig.\,\ref{fig:inversion_fit}. This is achieved for the absolute offsets for the three data sets as given in the top right of each panel. In Table\,\ref{tab:lit2}, we show the differences for the stellar parameters from the literature and from our best data inversion. Especially $R_*$ does differ significantly. Note, that $\Delta CB$, $i$, and $d\Omega$ were fixed for the inversion test. A histogram of the posterior distributions for all the inversion iterations of the five parameters of the 10 spots are shown in Fig.\,\ref{fig:spotdis}. We can see a clear clustering around longitudes 50 to 150 and 275 to 325\,$^{\circ}$, as well as co-latitudes ranging between 20 and 40\,$^{\circ}$. Spots show mostly 2 to 4\,$^{\circ}$ radius resulting in absolute spot filling factors of 2.5 to 3.5\,\%. Spot lifetimes seem to be mostly in-between 40 to 60\,d coinciding thereby with the results from the GP as shown in Fig.\,\ref{fig:QPC}.

\begin{table}
\caption{Differences of important stellar parameters from the literature and from our data inversion with {\tt starsim}} \label{tab:lit2}
\centering                         
\begin{tabular}{l c c}     
\hline\hline                 
\noalign{\smallskip}
\epseri & inv. & lit. \\
\hline
\noalign{\smallskip}
%effective temperature $T_{\rm eff}$ [K] & 5000 & 5076 \\ not important to mention here
spot temperature difference $\Delta T$ [K] & 500 & 1080 \\
convective blueshift difference CB [CB$_\odot$] & 0 & $\sim$0.3 \\ % this was set to 0
rotation period $P_{\rm rot}$ [day] & 11.33 & 11.2 \\
stellar radius $R_*$ [R$_{\odot}$] & 0.85 & 0.74 \\ 
stellar mass $M_*$  [M$_{\odot}$] & 0.85 & 0.82 \\   
%surface gravity $\log{\rm g}$ [dex] & 4.5 & 4.3 \\ % fixed before
inclination $i$ [deg]  & 90 & 60 \\ % this was set to 90
differential rotation $d\Omega$ [$d\Omega_{\odot}$] & 0 & 1.3 \\
\noalign{\smallskip} \hline               
\end{tabular}
\tablefoot{See Table\,\ref{tab:teststars} for details.}
\end{table}

\section{Tables of observational data}

% We could put Exptime, airmass, SNR here as well?
\begin{table*}
\caption{66 nightly binned time-series data points extracted with {\tt raccoon} from 204 HARPS spectra of $\varepsilon$\,Eri.} \label{table:HNdata}
\centering 
\scriptsize
\begin{tabular}{c c c c c}     
\hline\hline \noalign{\smallskip} 
BJD (d) & RV (m\,s$^{-1}$) &  BIS (m\,s$^{-1}$) & FWHM (m\,s$^{-1}$) & CON (\%) \\
\noalign{\smallskip} \hline \noalign{\smallskip}
2458761.86 & 16458.65$\pm$0.87 & 30.9$\pm$1.9 & 6473.8$\pm$5.7 & 40.694$\pm$0.029 \\
2458762.86 & 16451.20$\pm$0.79 & 35.4$\pm$1.8 & 6477.3$\pm$5.9 & 40.700$\pm$0.029 \\
2458763.87 & 16447.57$\pm$0.89 & 38.0$\pm$1.9 & 6473.5$\pm$6.1 & 40.747$\pm$0.031 \\
2458764.88 & 16444.59$\pm$0.87 & 36.5$\pm$1.8 & 6459.6$\pm$6.0 & 40.802$\pm$0.030 \\
2458765.85 & 16450.26$\pm$1.97 & 34.5$\pm$4.1 & 6452.2$\pm$6.0 & 40.846$\pm$0.030 \\
2458766.85 & 16453.40$\pm$0.90 & 30.5$\pm$1.7 & 6456.1$\pm$5.7 & 40.794$\pm$0.029 \\
2458767.71 & 16451.34$\pm$1.48 & 30.4$\pm$3.7 & 6457.0$\pm$6.1 & 40.796$\pm$0.029 \\
2458768.74 & 16448.66$\pm$1.47 & 34.7$\pm$3.0 & 6456.3$\pm$5.6 & 40.773$\pm$0.030 \\
2458773.86 & 16459.30$\pm$1.01 & 29.8$\pm$1.9 & 6481.9$\pm$5.5 & 40.662$\pm$0.029 \\
2458774.85 & 16450.86$\pm$0.90 & 36.3$\pm$2.4 & 6484.5$\pm$5.8 & 40.657$\pm$0.029 \\
2458775.83 & 16447.62$\pm$1.35 & 41.3$\pm$2.6 & 6474.1$\pm$6.1 & 40.698$\pm$0.031 \\
2458776.85 & 16449.16$\pm$1.71 & 38.4$\pm$3.3 & 6463.9$\pm$6.1 & 40.762$\pm$0.030 \\
2458777.86 & 16451.50$\pm$1.20 & 35.2$\pm$2.6 & 6463.2$\pm$5.8 & 40.746$\pm$0.029 \\
2458780.88 & 16445.40$\pm$3.00 & 34.0$\pm$6.2 & 6442.8$\pm$6.2 & 40.823$\pm$0.030 \\
2458781.64 & 16450.38$\pm$0.99 & 30.6$\pm$1.9 & 6449.9$\pm$5.6 & 40.798$\pm$0.028 \\
2458782.84 & 16453.43$\pm$1.32 & 27.8$\pm$3.4 & 6453.0$\pm$5.6 & 40.792$\pm$0.028 \\
2458783.85 & 16459.83$\pm$1.04 & 24.6$\pm$1.8 & 6464.0$\pm$5.5 & 40.732$\pm$0.028 \\
2458784.66 & 16462.15$\pm$1.18 & 23.5$\pm$2.4 & 6479.0$\pm$5.5 & 40.683$\pm$0.027 \\
2458785.65 & 16458.54$\pm$1.45 & 28.3$\pm$3.1 & 6487.6$\pm$5.5 & 40.661$\pm$0.027 \\
2458786.86 & 16453.31$\pm$1.36 & 36.6$\pm$2.6 & 6485.2$\pm$6.0 & 40.647$\pm$0.030 \\
2458791.67 & 16448.99$\pm$1.12 & 39.9$\pm$2.3 & 6454.3$\pm$6.0 & 40.782$\pm$0.030 \\
2458792.66 & 16452.07$\pm$1.19 & 34.0$\pm$2.5 & 6448.4$\pm$5.9 & 40.801$\pm$0.029 \\
2458793.67 & 16453.47$\pm$1.50 & 28.2$\pm$3.0 & 6447.6$\pm$5.6 & 40.808$\pm$0.028 \\
2458794.68 & 16455.97$\pm$1.52 & 25.0$\pm$3.2 & 6455.8$\pm$6.4 & 40.777$\pm$0.029 \\
2458795.65 & 16460.26$\pm$1.50 & 25.6$\pm$3.1 & 6472.3$\pm$5.7 & 40.724$\pm$0.029 \\
2458796.68 & 16458.50$\pm$1.41 & 29.0$\pm$2.7 & 6484.9$\pm$6.0 & 40.671$\pm$0.029 \\
2458797.64 & 16458.26$\pm$1.37 & 30.8$\pm$3.3 & 6482.9$\pm$5.9 & 40.672$\pm$0.029 \\
2458798.64 & 16460.87$\pm$1.50 & 29.1$\pm$3.0 & 6477.4$\pm$5.6 & 40.665$\pm$0.028 \\
2458799.68 & 16460.42$\pm$1.65 & 30.4$\pm$3.0 & 6484.9$\pm$5.5 & 40.619$\pm$0.028 \\
2458801.59 & 16447.92$\pm$0.98 & 46.8$\pm$2.0 & 6477.2$\pm$6.1 & 40.635$\pm$0.031 \\
2458802.57 & 16447.56$\pm$0.97 & 45.4$\pm$1.8 & 6460.4$\pm$6.1 & 40.719$\pm$0.031 \\
2458803.58 & 16450.37$\pm$1.04 & 38.6$\pm$2.3 & 6449.0$\pm$5.9 & 40.792$\pm$0.030 \\
2458804.63 & 16453.41$\pm$1.23 & 34.2$\pm$2.8 & 6446.3$\pm$5.8 & 40.816$\pm$0.030 \\
2458805.64 & 16455.08$\pm$0.96 & 28.0$\pm$1.7 & 6454.5$\pm$5.7 & 40.799$\pm$0.028 \\
2458806.63 & 16457.23$\pm$1.02 & 25.8$\pm$2.3 & 6463.3$\pm$5.6 & 40.770$\pm$0.029 \\
2458807.60 & 16456.94$\pm$0.81 & 28.2$\pm$1.6 & 6471.5$\pm$5.6 & 40.740$\pm$0.028 \\
2458808.56 & 16452.57$\pm$1.35 & 32.1$\pm$2.7 & 6470.5$\pm$5.8 & 40.737$\pm$0.029 \\
2458810.60 & 16464.52$\pm$1.33 & 22.6$\pm$2.4 & 6472.3$\pm$5.5 & 40.683$\pm$0.027 \\
2458811.61 & 16459.12$\pm$1.16 & 31.3$\pm$2.4 & 6484.8$\pm$5.5 & 40.617$\pm$0.028 \\
2458813.60 & 16447.59$\pm$1.12 & 46.6$\pm$2.3 & 6461.7$\pm$6.2 & 40.708$\pm$0.031 \\
2458818.79 & 16452.10$\pm$0.82 & 28.7$\pm$1.6 & 6469.7$\pm$5.5 & 40.754$\pm$0.028 \\
2458819.72 & 16450.28$\pm$0.96 & 31.8$\pm$1.8 & 6466.3$\pm$5.8 & 40.757$\pm$0.029 \\
2458826.59 & 16448.45$\pm$0.72 & 39.8$\pm$1.7 & 6448.1$\pm$5.8 & 40.784$\pm$0.030 \\
2458827.61 & 16455.48$\pm$1.24 & 31.0$\pm$2.3 & 6448.5$\pm$5.8 & 40.808$\pm$0.029 \\
2458828.59 & 16456.04$\pm$1.56 & 27.9$\pm$2.8 & 6454.4$\pm$5.7 & 40.810$\pm$0.029 \\
2458829.61 & 16455.77$\pm$1.15 & 29.3$\pm$2.6 & 6464.5$\pm$5.6 & 40.781$\pm$0.029 \\
2458830.58 & 16453.45$\pm$0.91 & 30.7$\pm$1.8 & 6467.7$\pm$5.7 & 40.768$\pm$0.029 \\
2458831.59 & 16455.78$\pm$1.06 & 26.5$\pm$1.8 & 6463.8$\pm$4.2 & 40.765$\pm$0.026 \\
2458832.57 & 16458.44$\pm$1.26 & 23.7$\pm$2.5 & 6469.6$\pm$5.4 & 40.708$\pm$0.028 \\
2458833.58 & 16459.41$\pm$1.07 & 26.1$\pm$2.3 & 6474.7$\pm$5.6 & 40.665$\pm$0.027 \\
2458834.58 & 16458.73$\pm$1.02 & 32.6$\pm$2.2 & 6476.5$\pm$5.4 & 40.635$\pm$0.027 \\
2458835.58 & 16453.54$\pm$0.78 & 40.8$\pm$1.7 & 6472.4$\pm$5.7 & 40.643$\pm$0.029 \\
2458836.60 & 16450.49$\pm$1.14 & 43.2$\pm$2.1 & 6453.3$\pm$6.0 & 40.729$\pm$0.030 \\
2458837.67 & 16452.31$\pm$1.09 & 35.2$\pm$2.5 & 6447.3$\pm$5.9 & 40.775$\pm$0.030 \\
2458838.68 & 16453.60$\pm$1.39 & 31.1$\pm$2.7 & 6447.1$\pm$5.6 & 40.788$\pm$0.029 \\
2458839.59 & 16455.40$\pm$0.93 & 31.3$\pm$1.8 & 6456.1$\pm$5.6 & 40.766$\pm$0.028 \\
2458840.69 & 16451.39$\pm$0.83 & 34.2$\pm$1.7 & 6462.9$\pm$5.9 & 40.753$\pm$0.030 \\
2458841.55 & 16449.60$\pm$1.25 & 36.0$\pm$2.7 & 6457.2$\pm$5.9 & 40.791$\pm$0.030 \\
2458842.55 & 16448.93$\pm$1.35 & 33.4$\pm$2.7 & 6451.5$\pm$5.9 & 40.822$\pm$0.030 \\
2458843.55 & 16451.20$\pm$1.34 & 28.9$\pm$2.6 & 6452.2$\pm$6.0 & 40.798$\pm$0.030 \\
2458844.58 & 16453.62$\pm$0.84 & 25.2$\pm$1.7 & 6455.4$\pm$5.6 & 40.754$\pm$0.028 \\
2458845.58 & 16457.42$\pm$1.40 & 27.6$\pm$2.9 & 6461.9$\pm$4.7 & 40.705$\pm$0.024 \\
2458846.58 & 16455.45$\pm$0.95 & 35.1$\pm$2.0 & 6468.1$\pm$5.6 & 40.670$\pm$0.028 \\
2458847.57 & 16451.02$\pm$1.11 & 38.5$\pm$2.1 & 6460.8$\pm$5.9 & 40.706$\pm$0.030 \\
2458848.57 & 16450.93$\pm$1.14 & 35.5$\pm$1.7 & 6453.5$\pm$4.5 & 40.748$\pm$0.023 \\
2458849.57 & 16455.84$\pm$0.90 & 30.9$\pm$2.1 & 6454.0$\pm$5.7 & 40.755$\pm$0.029 \\
\noalign{\smallskip} \hline              
\end{tabular}
\end{table*}

\begin{table*}
\caption{37 nightly binned time-series data extracted  with {\tt raccoon} from 75 CARMENES VIS spectra for AU\,Microscopii.} \label{table:CARMdata}
\centering
\begin{tabular}{c c c c c c c c}     
\hline \hline \noalign{\smallskip} 
BJD (d) & RV (m\,s$^{-1}$) &  BIS (m\,s$^{-1}$) & FWHM (m\,s$^{-1}$) & CON (\%) \\
\noalign{\smallskip} \hline \noalign{\smallskip}
2458678.57 & -5342.0$\pm$8.9 & 315.5$\pm$24.6 & 12687.3$\pm$55.4 & 10.092$\pm$0.035 \\
2458679.54$^*$ & -5003.1$\pm$6.2 & -87.2$\pm$15.6 & 12175.3$\pm$43.6 & 7.370$\pm$0.021 \\
2458680.54 & -5046.2$\pm$4.6 & -63.0$\pm$11.6 & 12285.4$\pm$39.7 & 9.843$\pm$0.025 \\
2458684.57 & -4889.9$\pm$12.0 & -244.6$\pm$30.3 & 12253.9$\pm$37.4 & 10.254$\pm$0.025 \\
2458686.55 & -4899.8$\pm$5.8 & -332.9$\pm$14.4 & 12385.5$\pm$44.7 & 10.124$\pm$0.029 \\
2458687.58 & -5168.7$\pm$8.7 & -6.0$\pm$22.0 & 12638.0$\pm$46.1 & 10.068$\pm$0.029 \\
2458688.58 & -5209.1$\pm$8.4 & 187.5$\pm$21.8 & 12275.3$\pm$46.7 & 10.042$\pm$0.030 \\
2458690.56 & -4965.3$\pm$4.6 & -201.1$\pm$11.3 & 12425.4$\pm$51.3 & 9.799$\pm$0.032 \\
2458691.51 & -4854.7$\pm$8.8 & -380.8$\pm$21.6 & 12653.4$\pm$58.2 & 10.266$\pm$0.037 \\
2458693.55 & -5200.6$\pm$3.9 & 21.0$\pm$9.8 & 12264.1$\pm$54.0 & 9.895$\pm$0.035 \\
2458694.60 & -4865.5$\pm$4.9 & -272.3$\pm$12.0 & 12334.8$\pm$55.4 & 10.209$\pm$0.036 \\
2458695.54 & -4994.6$\pm$12.0 & -98.0$\pm$30.6 & 12471.0$\pm$54.3 & 10.247$\pm$0.035 \\
2458696.52 & -4935.8$\pm$5.7 & -273.6$\pm$14.1 & 12794.4$\pm$72.0 & 9.978$\pm$0.044 \\
2458699.00 & -5043.8$\pm$3.4 & -104.1$\pm$8.5 & 12240.9$\pm$43.4 & 10.005$\pm$0.028 \\
2458700.48 & -5015.3$\pm$6.0 & -119.6$\pm$14.8 & 12303.3$\pm$54.1 & 10.211$\pm$0.036 \\
2458702.50 & -5220.2$\pm$16.9 & 67.3$\pm$42.6 & 12679.6$\pm$100.9 & 10.057$\pm$0.063 \\
2458704.49 & -4920.8$\pm$5.6 & -189.4$\pm$14.0 & 12219.0$\pm$54.7 & 10.124$\pm$0.036 \\
2458706.49 & -5126.3$\pm$9.7 & 7.2$\pm$24.9 & 12618.9$\pm$101.1 & 9.729$\pm$0.062 \\
2458706.50 & -5145.3$\pm$9.0 & -6.1$\pm$23.0 & 12638.4$\pm$104.0 & 9.672$\pm$0.063 \\
2458711.45 & -5214.7$\pm$4.7 & 171.6$\pm$12.5 & 12567.0$\pm$61.0 & 9.608$\pm$0.037 \\
2458712.45 & -5299.4$\pm$5.6 & 225.9$\pm$15.1 & 12481.2$\pm$63.8 & 9.888$\pm$0.040 \\
2458714.98 & -4939.9$\pm$4.1 & -133.8$\pm$10.2 & 12037.4$\pm$29.1 & 10.195$\pm$0.020 \\
2458718.45 & -5087.8$\pm$5.6 & 86.4$\pm$14.1 & 12125.5$\pm$39.6 & 10.303$\pm$0.027 \\
2458723.46 & -5047.8$\pm$5.3 & -41.7$\pm$13.2 & 12101.2$\pm$35.5 & 10.283$\pm$0.024 \\
2458724.42 & -5042.3$\pm$4.4 & -151.8$\pm$10.9 & 12349.0$\pm$46.3 & 9.914$\pm$0.030 \\
2458727.47 & -5239.3$\pm$6.9 & 21.5$\pm$17.3 & 12432.7$\pm$67.3 & 9.721$\pm$0.042 \\
2458742.39 & -5215.3$\pm$3.8 & 111.3$\pm$9.9 & 12342.7$\pm$63.9 & 9.736$\pm$0.040 \\
2458743.37 & -4897.4$\pm$6.3 & -267.0$\pm$15.4 & 12156.9$\pm$50.3 & 9.854$\pm$0.032 \\
2458744.35 & -5008.4$\pm$4.4 & -13.0$\pm$11.3 & 12055.4$\pm$41.0 & 10.105$\pm$0.027 \\
2458745.39 & -5055.2$\pm$6.7 & 68.0$\pm$17.3 & 12758.8$\pm$60.9 & 9.828$\pm$0.037 \\
2458755.40 & -5119.0$\pm$7.0 & -41.5$\pm$17.6 & 12372.5$\pm$56.2 & 9.948$\pm$0.036 \\
2458757.38 & -5110.3$\pm$11.5 & 116.6$\pm$29.7 & 12055.5$\pm$42.2 & 10.044$\pm$0.028 \\
2458759.37 & -4801.7$\pm$7.5 & -477.6$\pm$18.4 & 12297.0$\pm$45.6 & 10.384$\pm$0.031 \\
2458760.31 & -5119.5$\pm$4.5 & -92.3$\pm$11.2 & 12326.5$\pm$40.9 & 9.727$\pm$0.026 \\
2458761.35 & -5229.1$\pm$6.4 & 112.6$\pm$16.3 & 12412.7$\pm$45.4 & 10.218$\pm$0.030 \\
2458763.34 & -5031.1$\pm$5.2 & -50.2$\pm$13.0 & 12372.9$\pm$49.9 & 10.100$\pm$0.032 \\
2458765.33 & -5094.2$\pm$5.0 & -150.0$\pm$12.3 & 12478.2$\pm$51.0 & 9.990$\pm$0.032 \\
\noalign{\smallskip} \hline
\end{tabular}
\tablefoot{* marks the obvious data outlier for the CON measurement.}
\end{table*}

\end{appendix}

\end{document}